\documentclass[a4paper, 11pt]{article}
\usepackage[a4paper, inner=3.5cm, outer=3cm, top=2.5cm, bottom=2.5cm]{geometry}

\usepackage{amsmath,amssymb}   
\usepackage{bm}                
\usepackage{braket}            
\usepackage{graphicx}
\usepackage[sortcites=true]{biblatex}
\usepackage{hyperref}          
\hypersetup{
    colorlinks=true,
    citecolor=black,
    linkcolor=black,     
    urlcolor=cyan,
    }

\usepackage{xcolor}
\usepackage{xspace}
\newcommand{\integerpartition}{\textbf{\ref{app:integer-partition}}\xspace}
\newcommand{\hermiticity}{\textbf{\ref{app:hermiticity}}\xspace}
\newcommand{\gauge}{\textbf{\ref{app:gauge}}\xspace}
\newcommand{\vacuumstate}{\textbf{\ref{app:vacuum-state}}\xspace}
\newcommand{\renormalization}{\textbf{\ref{app:renormalization}}\xspace}
\newcommand{\oneoperatorterm}{\textbf{\ref{app:one-operator-term}}\xspace}
\newcommand{\quadraticform}{\textbf{\ref{app:quadratic-form}}\xspace}
\newcommand{\diagonalization}{\textbf{\ref{app:lswt-diagonalisation}}\xspace}
\newcommand{\analyticalmodels}{\textbf{\ref{app:analytical-models}}\xspace}

\newcommand{\thearticletitle}{Quantum Spin Wave Theory with Multi-Spin Interactions for Magnetic Materials Modeling}
\title{\thearticletitle}
\date{\today}

\usepackage{xspace}
\newcommand{\EqAllMightyHamiltonian}{\eqref{eq:all-mighty-hamiltonian}\xspace}
\newcommand{\EqSpinOperatorDeviation}{\eqref{eq:spin-operator-deviation}\xspace}
\newcommand{\EqRenormalizedAllMightyHamiltonian}{\eqref{eq:renormalized-all-mighty-hamiltonian}\xspace}
\newcommand{\EqRenormalizedParameters}{\eqref{eq:renormalized-parameters}\xspace}
\newcommand{\EqAllMightyHamiltonianUpToLswt}{\eqref{eq:all-mighty-hamiltonian-up-to-lswt}\xspace}
\newcommand{\EqOCoeff}{\eqref{eq:O-coeff}\xspace}
\newcommand{\EqVectorBosonsBeforeDiagonalization}{\eqref{eq:vector-bosons-before-diagonalization}\xspace}
\newcommand{\EqGrandDynamicalMatrixForm}{\eqref{eq:grand-dynamical-matrix-form}\xspace}
\newcommand{\EqBilinearPartBogoliubovHamiltonian}{\eqref{eq:bilinear-part-bogoliubov-hamiltonian}\xspace}
\newcommand{\EqGDMPlusMinusK}{\eqref{eq:GDM-plus-minus-k}\xspace}
\newcommand{\EqLinkEnergiesPlusMinusK}{\eqref{eq:link-energies-plus-minus-k}}
\newcommand{\EqLinkGPlusMinusK}{\eqref{eq:GDM-plus-minus-k}\xspace}
\newcommand{\EqMatrixToSerialDiagonalized}{\eqref{eq:matrix-to-serial-diagonalized}\xspace}
\newcommand{\EqModelOneOmegas}{\eqref{eq:model-1-omegas}\xspace}
\newcommand{\EqModelTwoOmegas}{\eqref{eq:model-2-omegas}\xspace}

\newcommand{\FigModelSA}{2a\xspace}
\newcommand{\FigModelSB}{2b\xspace}
\newcommand{\FigModelSAB}{2a,b\xspace}

\begin{document}

\begin{titlepage}
  \centering
  {\LARGE \thearticletitle\par}
  \vspace{0.5cm}

  {\large 
  Andrey Rybakov$^{*,1,2}$, 
  Marco Marino$^3$, 
  Yaroslav M. Blanter$^2$, 
  Eugenio Coronado$^1$,
  Jaime Ferrer$^{*,2,4,5}$}
  \vspace{0.5cm}

$^1$ Institute of Molecular Science (ICMol), University of Valencia, Valencia, 46980, Spain \par

$^2$ Kavli Institute of Nanoscience, Delft University of Technology, Lorentzweg 1, Delft, 2628, The Netherlands \par

$^3$ Department of Physics, TU Dortmund University, Dortmund, 44227, Germany \par

$^4$ Departamento de Física, Universidad de Oviedo, Oviedo, 33007, Spain \par

$^5$ Centro de Investigación en Nanomateriales y Nanotecnología, Universidad de Oviedo-CSIC, El Entrego, 33940, Spain\par

\vspace{1cm}
$^*$ andrey.rybakov@uv.es, ferrer@uniovi.es\\

\vspace{2cm}

\begin{abstract}
In the field of magnetism within condensed matter physics, magnetic systems are microscopically described by a spin Hamiltonian. Its parameterization is accessible through both ab initio calculations and fits to experimental data. Frequently, this is achieved via multi-stage calculations that either relate the parameters to observable material properties or map a Kohn-Sham Hamiltonian onto an effective spin model. In this realm, accurate and flexible methods are essential for simulating the physical properties of the latter. The present study focuses on one such method: spin wave theory. The approach is applied to a new formulation of the spin Hamiltonian, named polynomial localized spin model (PLSM), which includes terms of arbitrary order in the spin operators. This enables the direct application of linear spin wave theory (LSWT) to the analysis of the energy land-scape and spin-wave 
spectrum of arbitrary collinear and non-collinear phases in any realistic magnetic material. Moreover, the classical and quantum parts of the PLSM are cleanly separated. Finally, an open-source implementation of LSWT, named Magnopy, is presented. It interfaces with modern parameterization software, supports arbitrary collinear and non-collinear systems, and is designed for both scripting and standalone use.
\end{abstract}
\end{titlepage}

\section{\label{sec:introduction} Introduction}

Magnetic materials host collective excitations that are at the heart of the research field called "magnonics" or "magnon spintronics". These excitations are referred to as spin waves, and their quanta are termed magnons. Magnonics offers an alternative for information transfer and computing in the low-THz range, free from the Joule-heating problems faced by conventional approaches based on electron transport \cite{Chumak2015}. The majority of the experimental results have been achieved in ferrites, among which the yttrium iron garnet (YIG) holds a prominent position \cite{Serga2010}. However, recent progress in experimental research on two-dimensional (2D) van der Waals (vdW) materials \cite{Wang2022} has paved the way for novel fundamental studies and new practical applications of magnonics \cite{Flebus2024}.

\begin{figure}
  \centering
  \includegraphics[width=0.5\linewidth]{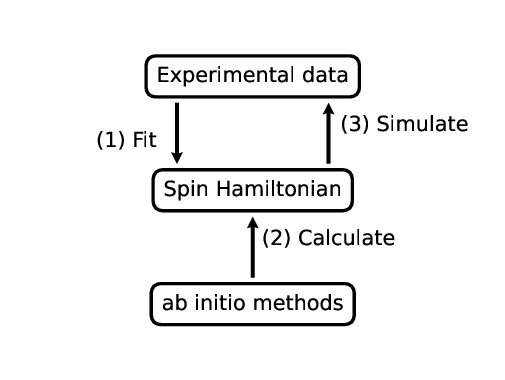}
  \caption{Scheme of a typical workflow that involves a spin Hamiltonian.}
  \label{fig:scheme}
\end{figure}

Moreover, the field of magnetism in theoretical condensed matter has entered an advanced era of its development \cite{Szilva2023}, in which quantitative predictions of physical phenomena and comparisons with experiments can and are expected to be performed. Modeling of the magnetic behavior of an arbitrary physical system, commonly starts with an effective \textit{spin Hamiltonian}, in which a set of \textit{interaction parameters} governs the system's magnetic state, evolution, and response. The interaction parameters are either treated as free parameters of the model, obtained from fitting the experimental data, or calculated \textit{ab initio} \cite{Liechtenstein1984,Drautz2004}. This situation can be represented with an abstract scheme as depicted in \textbf{Figure~\ref{fig:scheme}}. On one hand, a number of tools that can produce a spin Hamiltonian's parameterization (process 2 in Figure~\ref{fig:scheme}) are readily available for a wide array of researchers \cite{Korotin2015,He2021,MartnezCarracedo2023,Orlando2026}. On the other hand, a number of methods for simulating various magnetic properties across diverse types of parameterizations is implemented. Examples are classical or quantum Monte-Carlo methods \cite{landau2021guide,Evans2014}, atomistic spin dynamics (ASD) \cite{Evans2014,Skubic2008,Mller2019,sunny-arxiv}, or spin wave theory (SWT) \cite{sunny-arxiv,Toth2015}. These simulation tools can be used to either compute an observable property from the given parameterization of the spin Hamiltonian (process 3 in Figure~\ref{fig:scheme}) or, in reverse, to fit the experimental results and extract the parameterization of the Hamiltonian (process 1 in Figure~\ref{fig:scheme}).

However, the parameters of the spin Hamiltonian are not directly observable, meaning that an ultimate test for any theory — an experiment — can not define them unequivocally. In other words, the values of the parameters are as accurate as the underlying theory used to fit the experiment or in ab initio methods and as reliable as the implementation of that theory in a computational tool. Therefore, a mistake or misconception in any of those tools or theories can lead to the misinterpretation of the parameters and, consequently, of the physical phenomena. For example, imagine that the parameterization of the spin Hamiltonian is obtained from the fit of an experimental measurement via a theoretical method (for instance, via fitting of the inelastic neutron scattering data using the LSWT), and a mistake is hidden in the theoretical method that has been used. Then, when that parameterization is used as the starting point for a simulation of another observable (for instance, of the critical temperature via ASD), the error can lead to an over- or underestimation of that observable. Alternatively, one might attempt to parameterize the same system from ab initio, and the result of this calculation would not be compatible with the model fitted to experimental data. Moreover, as ab initio methods are not exact, such a mismatch can promote over-correction (e.~g., fine-tuning the Hubbard parameter in the DFT calculation, or fitting exchange parameters with an insufficient number of neighbor shells). Finally, the interaction parameters are subject to the ambiguity of the Hamiltonian's convention, which further complicates their comparison.

Motivated by the need for the accurate simulation of magnetic materials, in this work we revisit the SWT method, which is often employed for the simulation or the fitting (procedures (3) and (1) in Figure~\ref{fig:scheme}), with the goal of extending it for an arbitrary spin Hamiltonian with any number of spin operators. 
In doing so, we  highlight the applicability limits of SWT and discuss ambiguities that are present in the literature about LSWT implementation \cite{Toth2015} and affect the clarity of the fit from the experimental data \cite{Scheie2022}. 
Moreover, to facilitate application of the algebra for simulation of magnons in any real magnetic material, we introduce the open-source implementation of LSWT, \textit{Magnopy} \cite{footnote-magnopy}, which is convention-agnostic and can be readily used in tandem with both ab initio-based methods \cite{MartnezCarracedo2023, He2021} and simulation tools \cite{Skubic2008,Evans2014}. 
Finally, as LSWT can be performed analytically on compact spin Hamiltonians, we present a minimal set of well-defined models together with their analytical solutions. These models can be used to test the correctness of any numerical implementation of LSWT, independently of inaccuracies arising from the values of the interaction parameters, and we use them to benchmark Magnopy.

We organize this article into five sections, each dedicated to a clear objective. In the Section~\ref{sec:hamiltonian}, we introduce the model's formalism, which can be applied to a wide range of localized magnetic systems, and briefly describe SWT's place within this model. In the Section~\ref{sec:vacuum-state}, we define an a priori known vacuum state and relate it to the true quantum eigenstate of the spin Hamiltonian. The former serves as a ground level for the low-energy expansion. Then, in the Section~\ref{sec:classical-and-quantum}, we introduce the first approximation, which is specific to the LSWT, while we transform the Hamiltonian into its renormalized form. The latter highlights the assumptions introduced in the Section~\ref{sec:vacuum-state} and results in what can be regarded as Hamiltonian's classical and quantum parts. In the Section~\ref{sec:approximalte-bosonic}, we approximate the initial spin Hamiltonian with the bilinear boson Hamiltonian, and improve a diagonalization procedure of the latter. Finally, in the Section~\ref{sec:benchmarking}, we describe two analytically solvable models to test the correctness of numerical implementation of the LSWT. Moreover, we highlight the relevant portion of Magnopy's implementation in a short paragraph at the end of each section.

\section{\label{sec:hamiltonian} Hamiltonian}

We assume a set of magnetic sites placed within the unit cell of an arbitrary periodic lattice (one-, two-, or three-dimensional). The lattice is defined by the three non-coplanar lattice vectors $\bm{a}_1, \bm{a}_2, \bm{a}_3$ and comprises $N = N_1 \times N_2 \times N_3$ unit cells. The position of each unit cell within the lattice is described by the vector $\bm{r}_{\mu+\nu_m}$, and we always indicate unit cells by the indices $\mu$ and $\nu_m$. The unit cell indices denote a set of three integers, one per lattice vector. Each unit cell contains $M$ magnetic sites at positions $\bm{r}_{\alpha_m}$, and we always indicate magnetic sites by indices $\alpha_m$. Magnetic site indices denote a set of three real numbers from the half-open interval $[0,1)$. Indices of the unit cells can be added together to indicate addition of the radius-vector, for instance $\mu+\nu_m$ indicates the position $\bm{r}_{\mu} + \bm{r}_{\nu_m}$. The overall position of a magnetic site in the lattice is indicated by a pair of indices $(\mu+\nu_m,\alpha_m)$ that corresponds to the position $\bm{r}_{\mu+\nu_m}+\bm{r}_{\alpha_m}$, where we maintain the correspondence $\mu+\nu_m \leftrightarrow \alpha_m$. Finally, by notation, $\nu_1$ is always interpreted as $0$, i.~e. $\mu+\nu_1 \equiv \mu$. In the text, we assume $\hbar = 1$; thus, interaction parameters are understood to be given in units of energy.

Next, a three-dimensional spin operator $\bm{\hat{S}}_{\mu,\alpha}$, that fulfills the usual commutation relations of  angular momentum operator, is associated with each magnetic site. Note that domain of SWT's applicability is limited already here, as we assume that each spin is localized in real space and has a well-defined spin value ($S_{\alpha}$ for the site located at $\bm{r}_{\mu,\alpha}$). This limitation excludes, for instance, itinerant magnetism in metallic systems. The spin operator can always be expressed as
\begin{equation}
    \bm{\hat{S}}_{\mu, \alpha}
    =
    \hat{S}^{x_{\alpha}}_{\mu, \alpha}
    \bm{x}_{\alpha}
    +
    \hat{S}^{y_{\alpha}}_{\mu, \alpha}
    \bm{y}_{\alpha}
    +
    \hat{S}^{z_{\alpha}}_{\mu, \alpha}
    \bm{z}_{\alpha}, 
    \label{eq:spin-operator}
\end{equation}
where vector $\bm{z}_{\alpha} = \bm{R}_{\alpha}\bm{z}$ is a local direction axis for the magnetic site located at $\bm{r}_{\mu,\alpha}$, and $(\bm{x}_{\alpha}, \bm{y}_{\alpha}, \bm{z}_{\alpha})$ defines a local Cartesian reference frame for each site within the unit cell. Rotation matrix $\bm{R}_{\alpha}$ describes a rotation around the axis $\bm{z} \times \bm{z}_{\alpha}$ by an angle $\theta_{\alpha}$ subtended between $\bm{z}_{\alpha}$ and $\bm{z}$ in accordance with the right-hand rule. Note that all interaction parameters are always given in a global reference frame $(\bm{x}, \bm{y}, \bm{z})$. The operator $\hat{S}^{z_{\alpha}}_{\mu, \alpha}$ has $2S_{\alpha}+1$ eigenstates $\ket{S_{\mu,\alpha}^{z_{\alpha}} = -S_{\alpha}}, ..., \ket{S_{\mu,\alpha}^{z_{\alpha}} = S_{\alpha}}$, that span the local Hilbert space $\mathcal{L}_{\mu, \alpha}$  for each site. 
The corresponding eigenvalues are integer or half-integer numbers ranging from $-S_{\alpha}$ to $S_{\alpha}$.
Then, the Hilbert space of the whole magnetic system, $\mathcal{L}$, can be defined as a direct product of the local Hilbert spaces $\mathcal{L} = \otimes_{\mu,\alpha}\mathcal{L}_{\mu, \alpha}$. Note that every choice of a set of $M$ local reference frames fixes a basis in $\mathcal{L}$, but does not alter the Hilbert space.

At this point, we defined a set of magnetic sites that constitute the model of a physical system. We are now ready to define the Hamiltonian that captures the collective behavior of the set of magnetic sites, thereby modeling the behavior of a physical system. We choose to consider any system whose internal interactions and external effects can be well captured by the Hamiltonian expressed as a polynomial of the spin vectors or their components in the following generic form
\begin{equation}
    \mathcal{\hat{H}}
    =
    \sum_{n=1}^{\infty}
    C_n
    \sum_{\substack{\mu, \nu_2,\dots,\nu_n,\\ \alpha_1,\dots,\alpha_n,\\ i_1,\dots,i_n}}
    \Biggl(
        J^{i_1,\dots,i_n}_{\nu_2,\dots,\nu_n;\alpha_1,\dots,\alpha_n}
        \prod_{m=1}^n \hat{S}^{i_m}_{\mu+\nu_m,\alpha_m}
    \Biggr).
    \label{eq:all-mighty-hamiltonian}
\end{equation}

Here, $C_n$ are real constants that serve to embrace all the conventions of the spin Hamiltonian existing in the literature \cite{Ashcroft,simon2013oxford,Auerbach} that we are aware of, and the summation over the magnetic site indices includes all possible combinations of sites. The real $n$-dimensional tensors $J^{i_1,\dots,i_n}_{\nu_2,\dots,\nu_n;\alpha_1,\dots,\alpha_n}$ are the interaction parameters. This Hamiltonian captures all relevant interactions: the Zeeman coupling with the external magnetic field, the classical magnetic dipole-dipole interaction, the scalar Heisenberg exchange, 
the Dzyaloshinskii–Moriya interaction (DMI) \cite{Dzyaloshinsky1958,Moriya1960}, anisotropies of arbitrary order, biquadratic exchange interactions, that are found to be important in many 2D systems \cite{Kartsev2020,Ferriani2008,Mishra2021}, and other terms that do not have a universally accepted name yet \cite{Hoffmann2020}. Note that translational invariance of the Hamiltonian's parameters is reflected in their independence on the index $\mu$. We propose to call the Hamiltonian in Equation~\eqref{eq:all-mighty-hamiltonian}  a  \textit{polynomial localized spin model} (PLSM). We note that, by setting $n=2$, one can obtain from PLSM a generalized \textit{Heisenberg model} that includes arbitrary symmetric and asymmetric bilinear anisotropies in addition to the purely isotropic exchange. 

The distinction between the pair of spin operators belonging to two different magnetic sites or to the same magnetic site is fundamental. In the former case two spin operators commute, in the latter case they do not. Consequently the algebraic structure and physical nature of the terms in Equation~\eqref{eq:all-mighty-hamiltonian} depends on both: the total number of operators $n$ and on how those operators are distributed among distinct magnetic sites. This distinction leads to the classification of the terms of the Hamiltonian into \textit{equivalence classes}, that are identified by the total number of operators $n$ and by the integer partition of $n$ (see \integerpartition). Moreover, we identify a gauge invariance of the Hamiltonian, that is formulated within each equivalence class. The hermiticity of the Hamiltonian (see \hermiticity) and gauge fixing (see \gauge) lead to a restriction on the allowed form of the interaction parameter's tensors within each equivalence class. In this work we assume the Hamiltonian to be Hermitian and select a \textit{symmetric gauge} as defined in \gauge. For instance, for the terms with two spin operators hermiticity implies that $J^{i_1,i_2}_{0;\alpha_1,\alpha_1} = J^{i_2,i_1}_{0;\alpha_1,\alpha_1}$, while the symmetric gauge implies that $J^{i_1,i_2}_{\nu_2;\alpha_1,\alpha_2} = J^{i_2,i_1}_{-\nu_2;\alpha_2,\alpha_1}$ for terms with $(\nu_2, \alpha_2) \ne (0, \alpha_1)$. We note that in the special case of $n=2$ hermiticity and symmetric gauge combine into a single property, that is often written as $J^{i_1,i_2}_{\nu_2;\alpha_1,\alpha_2} = J^{i_2,i_1}_{-\nu_2;\alpha_2,\alpha_1}$. However, for $n \ge 3$ hermiticity and the choice of symmetric gauge produce two distinct groups of properties of interaction parameters. Hermiticity has a clear physical meaning for the Hamiltonian, while the gauge fixing is a convention choice.

With the system's model defined, we can proceed to consider the approach of SWT. The operators $\hat{S}^{z_{\alpha}}_{\mu,\alpha}$ do not commute with the Hamiltonian of Equation~\eqref{eq:all-mighty-hamiltonian} in general. Therefore, the task of finding its eigenstates in terms of linear combinations of the basis states from $\mathcal{L}$ is equivalent to diagonalization of a square matrix with dimensions proportional to $\sim const^{MN}$. Due to the exponential growth of the problem's dimensionality, its direct solution becomes challenging or plainly impossible when the number of magnetic sites is large (specifically $NM \gtrsim 30$). In this context, the SWT is an approximate method that reduces the problem's complexity. In particular, SWT resorts to obtaining the low-energy eigenstates of the system with respect to a local or global energy minimum of the classical part of the spin Hamiltonian (to be defined later), which is assumed to be known a priori. This stable state can be an approximation to the true ground state, a metastable state, or, luckily, the true ground state. In this work, we refer to it as \textit{vacuum state}, with this name being motivated by the bosonic representation of the spin operator introduced later in the text.

In Magnopy, we implement a ``SpinHamiltonian'' object that, at the time of this publication, handles full tensors of interaction parameters for each term of the Hamiltonian in Equation~\eqref{eq:all-mighty-hamiltonian} with $n \le 4$. This object provides an automatic conversion between any pair of conventions and serves as Magnopy's main interface for manipulating the magnetic system. It supports unit cells with an arbitrary number of sites and an arbitrary number of interaction parameters, with both quantities limited solely by computational resources.

\section{\label{sec:vacuum-state} Vacuum state}

In this section, we discuss the types of vacuum states considered in this work and supported in Magnopy. First, note that the choice of the fixed set of local reference frames $(\bm{x}_{\alpha}, \bm{y}_{\alpha}, \bm{z}_{\alpha})$ in the definition of the spin operator in Equation~\eqref{eq:spin-operator} is irrelevant for the exact eigenvalue problem, as the nature of its full solution does not depend on the choice of the basis in the Hilbert space $\mathcal{L}$. However, the variation of local reference frames is essential for approximate methods that search for the ground state, particularly for ASD or classical Monte-Carlo (cMC). To demonstrate it let us consider a special state in $\mathcal{L}$, defined as $\ket{0}\equiv\ket{S^{z_{\alpha}}_{\mu, \alpha} = S_{\alpha}, \forall\mu, \forall\alpha}$ and analyze the expectation value of PLSM in that state $NE^{vacuum} \equiv \bra{0}\mathcal{\hat{H}}\ket{0}$, which can be written as $E^{vacuum} = E^{(0)} + E^{corr}$, where
\begin{equation}
    E^{(0)}
    =
    \sum_{n=1}^{\infty}
    C_n
    \sum_{\substack{\nu_2,\dots,\nu_n,\\ \alpha_1,\dots,\alpha_n,\\ i_1,\dots,i_n}}
    \Biggl(
        J^{i_1,\dots,i_n}_{\nu_2,\dots,\nu_n;\alpha_1,\dots,\alpha_n}
        \prod_{m=1}^n S_{\alpha_m}z^{i_m}_{\alpha_m}
    \Biggr),
    \label{eq:classical-energy}
\end{equation}
is the classical energy of the system per unit cell and $E^{corr}$ is a quantum correction to that classical energy. The expression for $E^{corr}$ is convoluted and thus relegated to the \vacuumstate for $n \le 4$. The approach of ASD or cMC methods is meaningful when two conditions are satisfied: (i) the state $\ket{0}$ is an eigenstate of the Hamiltonian in Equation~\eqref{eq:all-mighty-hamiltonian} or is at least a good approximation to it, and (ii) $E^{corr} \ll E^{(0)}$. On the contrary, SWT relies only on condition (i), and LSWT gives an approximate value for $E^{corr}$. Unfortunately, condition (i) holds only for a subset of possible parameterizations of the PLSM. 

Before the next section, we reflect on the fact that we chose to define only $M$ instead of $NM$ local reference frames in Equation~\eqref{eq:spin-operator}. If the reference frames have the same translational symmetry as the underlying lattice, one can take advantage of the Bloch theorem in the solution of the quadratic bosonic Hamiltonian (to be defined later), which delivers a substantial simplification in periodic magnetic systems. If the magnetic order is incommensurate with the underlying periodic structure, then the unit cell can be chosen either to reproduce exactly or at least to approximate to a satisfactory degree the true magnetic supercell. We acknowledge that it is possible to introduce an inter-cell spiral order on top of the periodic magnetic structure \cite{Toth2015}. However, that technique unnecessarily restricts the values of linear and bilinear interaction parameters; in fact, it requires that the matrices of all bilinear interaction parameters commute with the spiral rotation matrices, and that the vectors of all linear parameters be directed along the rotation axis of the spiral. Fulfillment of these conditions is not guaranteed for an arbitrary parametrization and Umklapp-like terms may appear. The latter require a correct folding of the Hamiltonian from the full Brillouin zone into a fraction of it, which is equivalent to the construction of a supercell in real space. As we wish to consider an arbitrary parametrization, we do not use this technique here.

In Magnopy, we implemented the numerical minimization algorithm described by Ivanov~et~al. \cite{Ivanov2021} to find local minima of either $E^{(0)}$ or $E^{(0)} + E^{corr}$. Moreover, Magnopy can compute both terms of $E^{vacuum}$ and check whether $\ket{0}$ is an eigenstate of the Hamiltonian for any given set of direction vectors. The exact equations that are checked for the latter can be found in \vacuumstate.

\section{\label{sec:classical-and-quantum} Classical and quantum part of the Hamiltonian}

We assume in this section that a specific choice of the vacuum state $\ket{0}$ has been made, where a given 
set of $M$ local reference frames $(\bm{x}_{\alpha}, \bm{y}_{\alpha}, \bm{z}_{\alpha})$ has been picked, and we use $\ket{0}$ to rewrite the Hamiltonian and introduce the first approximation that is specific to LSWT. Let us express the spin operator from Equation~\eqref{eq:spin-operator} as
\begin{equation}
    \bm{\hat{S}}_{\mu,\alpha}
    =
    S_{\alpha}\bm{z}_{\alpha} + \bm{\hat{\Delta}}_{\mu,\alpha},
    \label{eq:spin-operator-deviation}
\end{equation}
where we call $\bm{\hat{\Delta}}_{\mu,\alpha}$ a \textit{spin deviation operator}. One can directly check that $\bm{\hat{\Delta}}_{\mu, \alpha}$ satisfies the same commutation relations as the spin operator. By inserting Equation~\eqref{eq:spin-operator-deviation} into Equation~\eqref{eq:all-mighty-hamiltonian} we can rewrite the Hamiltonian as
\begin{eqnarray}
    \mathcal{\hat{H}}
    &=&
    NE^{(0)}
    +
    \sum_{n=1}^{\infty}
    \sum_{\substack{\mu, \nu_2\dots,\nu_n,\\ \alpha_1,\dots,\alpha_n,\\ i_1,\dots,i_n}}
    \Biggl(
        \tilde{J}^{i_1,\dots,i_n}_{\nu_2,\dots,\nu_n;\alpha_1,\dots,\alpha_n}
        \prod_{m=1}^n \hat{\Delta}^{i_m}_{\mu+\nu_m,\alpha_m}
    \Biggr),
    \label{eq:renormalized-all-mighty-hamiltonian}
    \\
    \tilde{J}^{i_1,\dots,i_n}_{\nu_2,\dots,\nu_n;\alpha_1,\dots,\alpha_n}
    &=&
    \sum_{m=n}^{\infty}
    C_m
    \sum_{\substack{\nu_{n+1},\dots,\nu_m,\\ \alpha_{n+1},\dots,\alpha_m,\\ i_{n+1},\dots,i_m}}
    \Biggl(
        \sum_{p_{n,m}}
        J^{p_{n,m}(i_1,\dots,i_m)}_{p_{n,m}(\nu_2,\dots,\nu_m;\alpha_1,\dots,\alpha_m)}
        \prod_{l=n+1}^m S_{\alpha_l}z^{i_l}_{\alpha_l}
    \Biggr),
    \label{eq:renormalized-parameters}
\end{eqnarray}
where $E^{(0)}$ is the value of classical energy \textit{per unit cell} defined by Equation~\eqref{eq:classical-energy} and $\tilde{J}^{i_1,\dots,i_n}_{\nu_2,\dots,\nu_n;\alpha_1,\dots,\alpha_n}$ are the renormalized interaction parameters. Index $p_{n,m}$ denotes $m! / n! (m-n)!$ codependent $n$-combinations of the $m$ Cartesian and position indices, that appear after the expansion of the product of $m$ binomials of the form Equation~\eqref{eq:spin-operator-deviation}. We choose to absorb the convention constants $C_m$ inside the renormalized parameters for convenience. The Equation~\eqref{eq:renormalized-parameters} can be simplified to Equation~(S50) when all spin operators are located at different positions in real space or when $n,m \le 2$.  We refer the reader to the \renormalization for more detailed discussion and for the formal description of the sum over $p_{n,m}$.

The renormalized Hamiltonian in Equation~\eqref{eq:renormalized-all-mighty-hamiltonian} separates cleanly the classical and quantum parts of the initial Hamiltonian. ASD or cMC only consider the classical part (i.~e. $E^{(0)}$) and optimize it over the set of local directional vectors $\bm{z}_{\alpha}$ defined by Equation~\eqref{eq:spin-operator}. In contrast, SWT goes beyond the classical term: it uses an approximate representation of the spin deviation operator and delivers a set of eigenstates that describe the excitations above the vacuum state; it also produces an approximate expression for $E^{corr}$.  Several SWT representations of the spin deviation operator in Equation~\eqref{eq:spin-operator-deviation} are used in the literature. Examples are the Holstein-Primakoff (HP) \cite{Holstein1940}, the  Dyson-Maleev \cite{Dyson1956_1,Dyson1956_2,maleev1958scattering} or the less known, but exact Agranovich-Toschich representation \cite{agranovich1968collective,Goldhirsch1979}. Furthermore, the bosonic square root appearing in the HP representation can be expanded in terms of the parameter $\delta_{\mu,\alpha} = \braket{\hat{a}_{\mu,\alpha}^{\dagger}\hat{a}_{\mu,\alpha}}/2S_{\alpha}$, by using either the Taylor or the Newton methods \cite{Knig2021}. We choose in this article to employ the HP representation and use the Taylor expansion of the square root of the particle number operator under the approximation $\delta_{\mu,\alpha} \ll 1$. The latter implies that the spin deviation at each magnetic site is small relative to $2S_{\alpha}$; see references \cite{Vogl2020, Knig2021} for an extended discussion. In practice, this assumption limits the applicability of LSWT to sufficiently low temperatures, as compared to the transition temperature to the paramagnetic state \cite{Bloch1962}. By keeping the terms up to order zero in the parameter $\delta_{\mu,\alpha}$, we get the following approximate representation of the spin deviation operator \cite{Toth2015} in terms of bosonic operators, 
\begin{equation}
    \bm{\hat{\Delta}}_{\mu, \alpha}
    \approx
    \sqrt{\dfrac{S_{\alpha}}{2}}
    \Bigl(
    \overline{\bm{p}_{\alpha}}
    \hat{a}_{\mu,\alpha}
    +
    \bm{p}_{\alpha}
    \hat{a}^{\dagger}_{\mu,\alpha}
    \Bigr)
        -
    \bm{z}_{\alpha}
    \hat{a}^{\dagger}_{\mu, \alpha}
    \hat{a}_{\mu,\alpha},
    \label{eq:deviation-operator-HP-Taylor}
\end{equation}
where $\bm{p}_{\alpha} = \bm{x}_{\alpha} + i\bm{y}_{\alpha}$ are complex-valued vectors, the overline character indicates component-wise complex conjugation, $\hat{a}_{\mu,\alpha}$ and $\hat{a}_{\mu,\alpha}^{\dagger}$ are HP bosonic creation and annihilation operators associated with the 
magnetic site located at $\bm{r}_{\mu,\alpha}$. The HP representation of the spin operators allows one to recast the spin Hamiltonian that acts onto the Hilbert space $\mathcal{L}$ as a bosonic Hamiltonian acting on a Fock space. The states in the latter are defined by the occupation numbers of the bosonic quasi-particle states called \textit{magnons}. Consequently, the diagonalization of the spin Hamiltonian is reformulated as the simpler problem of the diagonalization of a bosonic Hamiltonian.

In Magnopy, we implement the calculation of the renormalized interaction parameters for any given set of direction vectors for $n,m \le 4$ with the exact formulas being provided in \renormalization.

\section{\label{sec:approximalte-bosonic} Approximate bosonic Hamiltonian}

Finally, all necessary pieces are in place to discuss the bosonic Hamiltonian. We insert the approximate bosonic representation of Equation~\eqref{eq:deviation-operator-HP-Taylor} into the Hamiltonian in Equation~\eqref{eq:renormalized-all-mighty-hamiltonian}. The resulting bosonic Hamiltonian is found to include not only bilinear but also biquadratic and higher \cite{footnote-high-boson-terms} boson terms that describe mutual boson interactions. LSWT keeps only the bilinear terms and discards biquadratic and higher order ones (i.~e. ignores all boson interactions). The LSWT ansatz cannot be justified for an arbitrary parametrization of the Hamiltonian in Equation~\eqref{eq:all-mighty-hamiltonian}. However, for many of those parameterizations, LSWT can be shown to be a good approximation. For example, correlations between spin deviations on different sites were discarded by Holstein and Primakoff \cite{Holstein1940} for a ferromagnet, leading to a bilinear Hamiltonian. Therefore, while presenting the LSWT algebra next, we encourage the reader and user of either Magnopy or of any other implementation of LSWT to be conscious and cautious about its applicability for each specific parameterization of PLSM.

From this point on, under all the approximations that have been described above, we discuss the quadratic part of the bosonic Hamiltonian. Here, the usefulness of the renormalized form of the Hamiltonian becomes apparent due to the fact that only the terms $n=1,\,2$ in the sum over $n$ in Equation~\eqref{eq:renormalized-all-mighty-hamiltonian} produce one- or two-operator terms. By applying the common unitary Fourier transform to them
\begin{subequations}
\begin{eqnarray}
    \hat{a}_{\mu,\alpha}
    &=&
    \dfrac{1}{\sqrt{N}}
    \sum_{\bm{k}}
    e^{i\bm{k}\cdot\bm{r}_{\mu}}
    \hat{a}_{\alpha}(\bm{k}),
    \\ 
    \hat{a}_{\mu,\alpha}^{\dagger}
    &=&
    \dfrac{1}{\sqrt{N}}
    \sum_{\bm{k}}
    e^{-i\bm{k}\cdot\bm{r}_{\mu}}
    \hat{a}_{\alpha}^{\dagger}(\bm{k}),
    \label{eq:bosonic-real-reciprocal}
\end{eqnarray}
\end{subequations}
where the reciprocal vectors $\bm{k}$ assume $N$ values distributed across the first Brillouin zone,
we arrive at the LSWT approximation of the Hamiltonian in Equation~\eqref{eq:all-mighty-hamiltonian}
\begin{subequations}
\begin{eqnarray}
    \mathcal{\hat{H}}
    &\approx&
    N\Bigl(E^{(0)} + E^{(2)}\Bigr)
    +
    \sqrt{N}
    \sum_{\alpha_1}
    \Bigl(
    O_{\alpha_1}
    \hat{a}_{\alpha_1}(\bm{0})
    +
    \overline{O_{\alpha_1}}
    \hat{a}^{\dagger}_{\alpha_1}(\bm{0})
    \Bigr)
    +\nonumber\\&&+
    \sum_{\bm{k}}
    \bm{\mathcal{\hat{A}}}^{\dagger}(\bm{k})
    \bm{\mathcal{D}}(\bm{k})
    \bm{\mathcal{\hat{A}}}(\bm{k}),
    \label{eq:all-mighty-hamiltonian-up-to-lswt}
    \\
    E^{(2)}
    &=&
    \dfrac{1}{2}
    \sum_{\alpha_1, i_1}
    z_{\alpha_1}^{i_1}
    \tilde{J}_{\alpha_1}^{i_1},
    \label{eq:energy-2}
    \\
    O_{\alpha_1}
    &=&
    \sum_{i_1}
    \sqrt{\dfrac{S_{\alpha_1}}{2}}
    \overline{p_{\alpha_1}^{i_1}}
    \tilde{J}_{\alpha_1}^{i_1},
    \label{eq:O-coeff}
\end{eqnarray}
\end{subequations}
where $E^{(2)}$ is a correction to the classical vacuum state energy that arises from LSWT (i.~e. it is the first term in $E^{corr}$) and $O_{\alpha_1}$ are the coefficients of the one-operator terms. The amplitude of one-operator terms is proportional to $\sqrt{N}$, therefore if $O_{\alpha_1} \neq 0$ then the vacuum state $\ket{0}$ of the Hamiltonian in Equation~\eqref{eq:all-mighty-hamiltonian} will be destroyed by the spontaneous creation of the bosonic excitation. Fortunately, if the ground state is an extremum of the classical energy $E^{(0)}$, then $O_{\alpha_1} = 0$ (for the proof of that statement we refer the reader to the \oneoperatorterm. Therefore, the condition $O_{\alpha_1} = 0$ can be used to find the classical ground state of PLSM \cite{YungLi1964}. The last term in Equation~\eqref{eq:all-mighty-hamiltonian-up-to-lswt} is the Bogoliubov boson Hamiltonian, where
\begin{equation}
    \bm{\mathcal{\hat{A}}}(\bm{k})
    =
    \Bigl(\hat{a}_{1}(\bm{k}), \dots, \hat{a}_{M}(\bm{k}), \hat{a}_{1}^{\dagger}(-\bm{k}), \dots, \hat{a}_{M}^{\dagger}(-\bm{k})\Bigr)^T
    \label{eq:vector-bosons-before-diagonalization}
\end{equation}
and the $2M\times 2M$ grand-dynamical matrix $\bm{\mathcal{D}}(\bm{k})$ has the well-known form \cite{Toth2015,Ohashi2020}
\begin{equation}
    \bm{\mathcal{D}}(\bm{k})
    =
    \begin{pmatrix}
        \bm{D_1}(\bm{k}) & \bm{D_2}(\bm{k}) \\
        \overline{\bm{D_2}(-\bm{k})} & \overline{\bm{D_1}(-\bm{k})} \\
    \end{pmatrix}.
    \label{eq:grand-dynamical-matrix-form}
\end{equation}

The expressions for the $M\times M$ matrices $\bm{D_1}(\bm{k})$ and $\bm{D_2}(\bm{k})$ arising from the linear and bilinear spin terms of the Hamiltonian in Equation~\eqref{eq:all-mighty-hamiltonian} have been reported before \cite{Toth2015} and obey $\bm{D_1}(\bm{k})=\bm{D_1}^\dagger(\bm{k})$, 
$\bm{D_2}^\dagger(\bm{k})=\overline{\bm{D_2}(-\bm{k})}$. They are modified only slightly in the case of the PLSM and we provide them in the \quadraticform. The last step of the solution consists of diagonalizing the Bogoliubov boson Hamiltonian and finding a new set of $M$ bosonic operators $\hat{b}^{\dagger}_{\beta}(\bm{k}), \hat{b}_{\beta}(\bm{k})$ that create/destroy its eigenstates. This task is achieved via a Bogoliubov transformation which is written as $\bm{\mathcal{\hat{B}}}(\bm{k}) = \bm{\mathcal{G}}(\bm{k})\bm{\mathcal{\hat{A}}}(\bm{k})$, where $\bm{\mathcal{\hat{B}}}(\bm{k})$ is a $2M$-component column vector comprising the new collective bosonic operators, and is defined similarly to $\bm{\mathcal{\hat{A}}}(\bm{k})$ in Equation~\eqref{eq:vector-bosons-before-diagonalization}; $\bm{\mathcal{G}}(\bm{k})$ is a para-unitary $2M\times 2M$ transformation matrix. The bilinear part of the Hamiltonian becomes
\begin{equation}
    \mathcal{\hat{H}}^{B}
    =
    \sum_{\bm{k}}
    \bm{\mathcal{\hat{A}}}^{\dagger}(\bm{k})
    \bm{\mathcal{D}}(\bm{k})
    \bm{\mathcal{\hat{A}}}(\bm{k})
    = 
    \sum_{\bm{k}}
    \bm{\mathcal{\hat{B}}}^{\dagger}(\bm{k})
    \bm{\mathcal{E}}(\bm{k})
    \bm{\mathcal{\hat{B}}}(\bm{k}),
    \label{eq:bilinear-part-bogoliubov-hamiltonian}
\end{equation}
where $\bm{\mathcal{E}}(\bm{k})$ is a $2M\times 2M$ diagonal matrix, whose elements are all positive and arranged as described by Colpa \cite{Colpa1978}. However, they are not doubly degenerate, as incorrectly stated by Toth and Lake \cite{Toth2015}; a counterexample is given by the Model 1 defined later in this manuscript. Moreover, the eigenenergy of each normal magnon mode $\hat{b}_{\beta}(\bm{k})$ enters two separate matrices: half in $\bm{\mathcal{E}}(\bm{k})$ and half in $\bm{\mathcal{E}}(-\bm{k})$. In fact, that separation can be regarded as the eigenenergies of "particle" and "hole" excitations \cite{Ohashi2020}. However, as they both describe the same physical excitations \cite{Ohashi2020}, they should be recombined to obtain the full solution. Fortunately, the grand-dynamical matrices for positive and negative $\bm{k}$ are connected via an involutory matrix $\bm{\mathcal{I}}$ as 
\begin{equation}
    \bm{\mathcal{D}}(\bm{k}) = \bm{\mathcal{I}}\overline{\bm{\mathcal{D}}(-\bm{k})}\bm{\mathcal{I}},
    \qquad
    \bm{\mathcal{I}}
    =
    \sigma_1 \otimes \bm{I}
    =
    \begin{pmatrix}
        \bm{0} & \bm{I} \\
        \bm{I} & \bm{0}
    \end{pmatrix},
    \label{eq:GDM-plus-minus-k}
\end{equation}
where $\sigma_1$ is the first Pauli matrix, $\bm{I}$ is an $M\times M$ identity matrix, and $\bm{0}$ is an $M\times M$ matrix with all elements equal to zero. The Equation~\eqref{eq:GDM-plus-minus-k} is a manifestation of the particle-hole symmetry of the Bogoliubov boson Hamiltonian \cite{Ohashi2020}. Based on the results by Colpa \cite{Colpa1978} and on Equation~\eqref{eq:GDM-plus-minus-k}, we show (see \diagonalization) that the order of the eigenvalues can be chosen such that the diagonal elements of the $\bm{\mathcal{E}}(\bm{k})$ are ordered as
\begin{subequations}
\begin{equation}
    \tilde{\omega}_1(\bm{k}),
    \dots,
    \tilde{\omega}_M(\bm{k}),
    \tilde{\omega}_{M+1}(\bm{k}),
    \dots,
    \tilde{\omega}_{2M}(\bm{k})
\end{equation}
and the diagonal elements of the $\bm{\mathcal{E}}(-\bm{k})$ are ordered as
\begin{equation}
    \tilde{\omega}_{M+1}(\bm{k}),
    \dots,
    \tilde{\omega}_{2M}(\bm{k}),
    \tilde{\omega}_1(\bm{k}),
    \dots,
    \tilde{\omega}_M(\bm{k}).
\end{equation}
\label{eq:link-energies-plus-minus-k}
\end{subequations}

Moreover, we demonstrate that (see \diagonalization)
\begin{equation}
    \bm{\mathcal{G}}(\bm{k}) = \bm{\mathcal{I}}\overline{\bm{\mathcal{G}}(-\bm{k})}\bm{\mathcal{I}}.
    \label{eq:link-G-plus-minus-k}
\end{equation}

Therefore, the diagonalization of $\bm{\mathcal{D}}(\bm{k})$ for any given $\bm{k}$ defines transformation matrices and eigenenergies for $-\bm{k}$ at no additional computational cost. After recombination of the terms with $\pm\bm{k}$ the Hamiltonian becomes
\begin{subequations}
\begin{eqnarray}
    \mathcal{\hat{H}}^{B}
    &=&
    \sum_{\bm{k}}
    \sum_{\beta}
    \omega_{\beta}(\bm{k})
    \Biggl(
        \hat{b}^{\dagger}_{\beta}(\bm{k})
        \hat{b}_{\beta}(\bm{k})
        +
        \dfrac{1}{2}
    \Biggr),
    \label{eq:matrix-to-serial-diagonalized}
    \\
    \omega_{\beta}(\bm{k})
    &=&
    2\tilde{\omega}_{\beta}(\bm{k}),
    \label{eq:matrix-to-serial-omegas}
\end{eqnarray}
\end{subequations}
where the mode index $\beta$ runs from $1$ to $M$ (we refer the reader to the \diagonalization for more details). We note that sometimes the Equation~\eqref{eq:all-mighty-hamiltonian-up-to-lswt} is rewritten with $\bm{k} > 0$ under the sum and the magnon frequencies are then defined as $\omega_{\beta}(\bm{k}) = \tilde{\omega}_{\beta}(\bm{k})$ and $\omega_{\beta}(-\bm{k}) = \tilde{\omega}_{M+\beta}(\bm{k})$. However, as in that case the grand-dynamical matrix would be modified from $\bm{\mathcal{D}}(\bm{k})$ to $2\bm{\mathcal{D}}(\bm{k})$, the values of $\omega_{\beta}(\bm{k})$ do not change and the end result is identical to the Equation~\eqref{eq:matrix-to-serial-diagonalized}.

All in all, the initial spin Hamiltonian~\eqref{eq:all-mighty-hamiltonian} diagonalized at the level of LSWT, assuming the vacuum state $\ket{0}$, is
\begin{equation}
    \mathcal{\hat{H}}
    \approx
    N
    \left(
        E^{(0)} 
        + 
        E^{(2)}
    \right)
    +
    \sum_{\bm{k}}
    \sum_{\beta}
    \omega_{\beta}(\bm{k})
    \left(
        \hat{b}^{\dagger}_{\beta}(\bm{k})
        \hat{b}_{\beta}(\bm{k})
        +
        \dfrac{1}{2}
    \right),
    \label{eq:final-hamiltonian}
\end{equation}
where the part of $E^{corr}$ that originates from LSWT can be identified as 
\begin{equation}
    E^{corr}
    =
    \underbrace{
    E^{(2)}
    +
    \dfrac{1}{2N}
    \sum_{\bm{k}}
    \sum_{\beta}
    \omega_{\beta}(\bm{k})}_{\text{LSWT}}
    +
    \ldots
    \label{eq:lswt-E-corr}
\end{equation}

In Magnopy, we implemented the calculation of $E^{(2)}$ and $O_{\alpha}$ for any given set of local direction vectors. The latter can be used to validate the stability of the vacuum state. Moreover, we implemented the numerical Bogoliubov transformation as described by Colpa in Section 3, Remark 1 \cite{Colpa1978}. This transformation yields the {\it magnon dispersion relation} $\omega_{\beta}(\bm{k})$ and transformation matrices $\bm{\mathcal{G}}(\bm{k})$. Note that the value of $E^{(2)}$ is of little importance as Magnopy is capable of computing the value of $E^{corr}$ directly, without the approximations tied to the LSWT.

\section{\label{sec:benchmarking} Benchmarking}

As more and more methods are available for the ab initio estimation of the interaction parameters \cite{MartnezCarracedo2023,He2021,Orlando2026,Pashov2020,Xiang2013}, one is tempted to illustrate the correctness of the LSWT implementation by computing the magnon dispersion of a particular material. However, such a calculation would depend not only on the quality of the LSWT implementation but also on the quality of the used interaction parameters. Therefore, we advocate using a set of well-defined \textit{analytical} magnon problems and, first of all, testing a numerical implementation against them. This test is enabled by the fact that the Bogoliubov transformation can be performed analytically in cases where the unit cell contains only a few magnetic sites.

\begin{figure}
  \includegraphics[width=\linewidth]{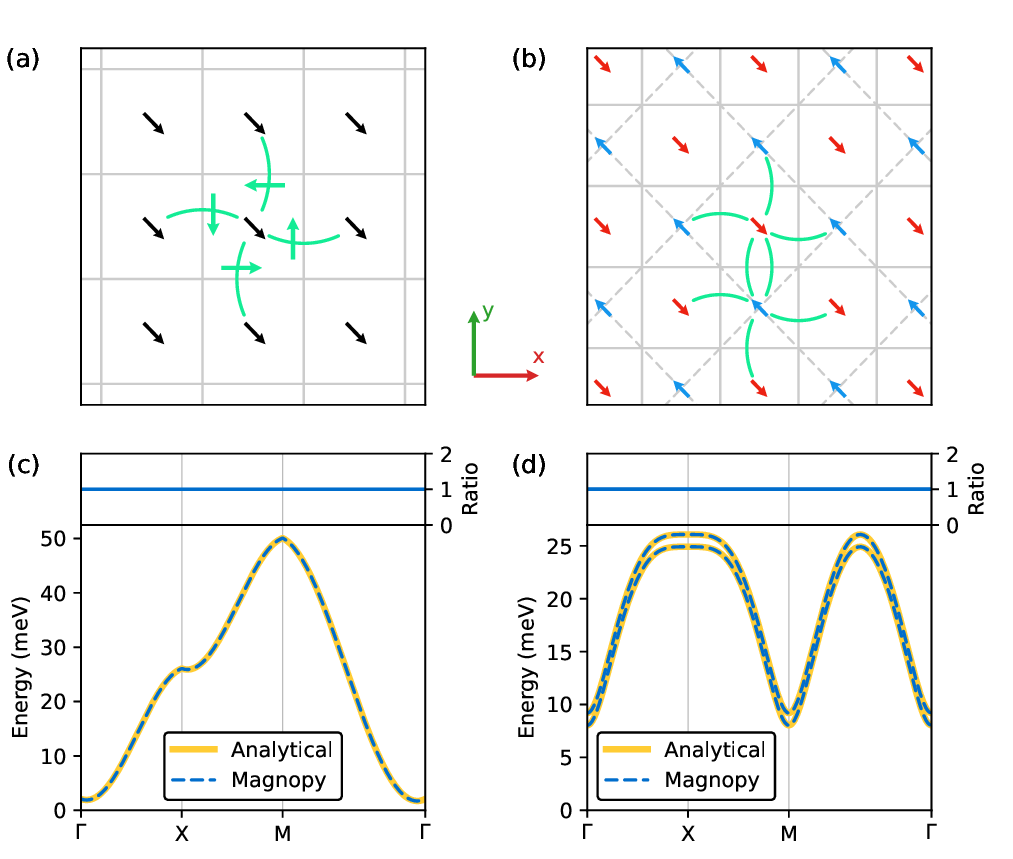}
  \caption{Two analytical models that test LSWT. (a) A ferromagnet on a square lattice. The unit cells of the lattice are indicated by gray solid lines. The directional vectors $\bm{z}_{\alpha}$ that define the ferromagnetic vacuum state are indicated by black arrows. Curved green lines indicate the four nearest neighbors of the primitive unit of the lattice. The directions of the DMI vectors are indicated by green arrows overlaid on top of each bond. The magnetic easy axis and the external magnetic field are collinear with the directional vectors $\bm{z}_{\alpha}$; (b) An antiferromagnet on the same square lattice. The unit cells of the crystal lattice are indicated by solid gray lines, and the unit cells of the magnetic lattice are twice larger and are indicated by dashed gray lines. Curved green lines indicate the eight nearest neighbors (four per site) of the primitive unit of the  magnetic lattice. The direction of the magnetic easy axis and of the external magnetic field is the same as in (a); (c, d) Magnon dispersion from the analytical solution (solid yellow lines) and from Magnopy's numerical solution (dashed blue lines) for the (c) ferromagnet, (d) antiferromagnet. The value of the parameters used in the calculations are $|J| = 2$~meV, $|B| = 5$~T, $A = 0.5$~meV, $|D| = 0.5$~meV, $S = 3 / 2$. The Ratio values in the top panels are defined as $\omega^{numerical}_{\beta}(\bm{k}) / \omega^{analytical}_{\beta}(\bm{k})$ and are computed individually for each mode and each $\bm{k}$. Global reference frame common for both models is depicted between the panels (a) and (b). For more technical details, we refer the reader to the \analyticalmodels.}
  \label{fig:models}
\end{figure}

In that spirit, we propose two examples of a generalized Heisenberg model of a 2D magnet, that comprise an absolute minimum that should be reproduced by any LSWT solver: a ferromagnet (Model 1) and an antiferromagnet (Model 2) on a square lattice. The two models are intentionally formulated for a 2D lattice (with 3D spins) to avoid potential confusion in the definition of the unit cell. The Hamiltonian for a ferromagnet includes isotropic exchange ($J$), a magnetic easy axis anisotropy term ($A$), a Zeeman term aligned with the easy axis ($B$), and a small antisymmetric DM interaction ($D$). The Hamiltonian for the antiferromagnet includes isotropic exchange ($J$), a magnetic easy-axis anisotropy term ($A$), and a Zeeman term directed along the easy axis ($B$). The models are schematically represented in \textbf{Figures~\ref{fig:models}a,b}, and we refer the reader to the \analyticalmodels for the exhaustive formal definition of both models. The analytical solution for Model 1 is

\begin{eqnarray}
    \omega(\bm{k})
    &=&
    g\mu_B B
    + 
    2SA
    -
    4SJ\Bigl(2 - \cos(\bm{k}\cdot\bm{a}_1) - \cos(\bm{k}\cdot\bm{a}_2)\Bigr)
    -\nonumber\\&&-
    2\sqrt{2}SD\Bigl(\sin(\bm{k}\cdot\bm{a}_1) + \sin(\bm{k}\cdot\bm{a}_2)\Bigr).
    \label{eq:model-1-omegas}
\end{eqnarray}

Model 2 has two normal modes, and its analytical solution is
\begin{subequations}
\begin{eqnarray}
    \omega_{1,2}(\bm{k})
    &=&
    \pm g\mu_B B
    +
    8SJ\sqrt{\Bigl(1 + \dfrac{A}{4J}\Bigr)^2 - \gamma(\bm{k}) \overline{\gamma(\bm{k})}},
    \\
    \gamma(\bm{k}) \overline{\gamma(\bm{k})}
    &=&
    \cos^2\left(\dfrac{\bm{k}\cdot\bm{a}_1}{2}\right)
    \cos^2\left(\dfrac{\bm{k}\cdot\bm{a}_2}{2}\right).
\end{eqnarray}
\label{eq:model-2-omegas}
\end{subequations}

The LSWT implementation in Magnopy shows a perfect agreement with the analytical solutions of both models, as illustrated in \textbf{Figures~\ref{fig:models}c,d}.

\section{\label{sec:conclusions} Conclusion}

We have proposed in this work a generic localized spin model that we have called Polynomial Localized Spin Model or PLSM in short. The Hamiltonian contains an arbitrary amount of spin operators ($n$) that interact via full interaction tensors. We have obtained exact expressions for the full quantum correction to the classical energy of the vacuum state for $n \le 4$, and have introduced a renormalization procedure that cleanly separates its classical and quantum parts. Then, we have computed all terms of the LSWT Hamiltonian that stem from the $n \le 4$ terms of the Equation~\eqref{eq:all-mighty-hamiltonian}. Note that the inclusion of the terms with $n > 4$ is possible by computing their contributions to the renormalized parameters $\tilde{J}_{\alpha_1}^{i_1}$ and $\tilde{J}_{\nu_2;\alpha_1,\alpha_2}^{i_1,i_2}$ and by using the equations presented in this work without any further modifications. The validity of LSWT for a given PLSM is subject to the following three requisites:
\begin{itemize}
    \item The vacuum state $\ket{\text{0}}$ has to be (close to) an eigenstate of the spin Hamiltonian in Equation~\eqref{eq:all-mighty-hamiltonian};
    \item The number of bosonic excitations on each site has to be small compared with the value of $2S_{\alpha}$ for that site ($\delta_{\mu,\alpha} << 1$);
    \item If higher than bilinear terms are left after the above two approximations, then the approximate quadratic form of the Hamiltonian in Equation~\eqref{eq:all-mighty-hamiltonian-up-to-lswt} has to be justified for the given parametrization of PLSM.
\end{itemize}
Moreover, we have demonstrated that diagonalizing of the LSWT Bogoliubov Hamiltonian for a single $\bm{k}$ provides the solution for both $\pm\bm{k}$ at no additional computational cost, thus by carefully sorting the k-points one can reduce the computational effort by a factor of $2$, for instance in the integration of arbitrary functions of magnon dispersion over the Brillouin zone.

In addition, we have proposed a minimal set of analytical examples that any LSWT numerical solver must reproduce, and we call for a collaborative effort among developers of magnon-related software to formulate a comprehensive set of standard problems.

Finally, we have introduced a numerical implementation of the manuscript's algebra, that we have named Magnopy. The code's capabilities include: (i) Finding the local minima of both $E^{(0)}$ and $E^{(0)}+E^{corr}$; (ii) Evaluating of all the coefficients that appear in the renormalized spin Hamiltonian and in the bosonic Hamiltonian, in particular $E^{(0)}$ (Equation~\eqref{eq:classical-energy}), $E^{(2)}$ (Equation~\eqref{eq:energy-2}), $E^{corr}$ (Equation~(S30)), $O_{\alpha_1}$ (Equation~\eqref{eq:O-coeff}), the renormalized interaction parameters $\tilde{J}^{i_1,\dots,i_{n\le 4}}_{\nu_2,\dots,\nu_{n\le 4};\alpha_1,\dots,\alpha_{n\le 4}}$ (Equation~\eqref{eq:renormalized-parameters} and Equation~(S51)~--~(S54)), the magnon dispersion relation $\omega_{\beta}(\bm{k})$; (iii) Checking some of the assumptions that validate the choice of a vacuum state, in particular whether $\ket{0}$ is an eigenstate of the spin Hamiltonian (Equation~(S34)~--~(S44)). Note that at the time of this publication, Magnopy supports all possible terms in the spin Hamiltonian in Equation~\eqref{eq:all-mighty-hamiltonian} up to $n = 4$ together with full 
interaction tensor parameters. Support for higher-order terms ($n>4$) is planned for future releases, as the algebraic formalism presented here naturally extends to more complex interactions. Magnopy is designed to operate both as a library for scripting and as a standalone program that can be executed from the command line. In the latter case, it outputs human-readable text to the console that explains the course of calculations in details and saves various numerical and visual data into separate files. Magnopy supports the definition of an arbitrary spin Hamiltonian (in any convention) and is interfaced with TB2J \cite{He2021} and GROGU \cite{MartnezCarracedo2023}, which can compute the interaction parameters from ab initio simulations. Magnopy is written in the Python programming language and depends on NumPy \cite{Harris2020} for manipulations with arrays and on spglib \cite{Togo2024} and Wulfric \cite{wulfirc-docs} for the crystal symmetry analysis and automatic generation of high-symmetry paths and k-points as described by Setyawan and Curtarolo (SC) \cite{Setyawan2010} or by Hinuma~et~al. (HPKOT) \cite{Hinuma2017} with HPKOT being a default option.

\medskip
\textbf{Data Availability}

The source code of Magnopy is openly available in GitHub at \url{https://github.com/magnopy/magnopy}.

\medskip
\textbf{Acknowledgements} \par 

A.~R. thanks Levente Rózsa for the fruitful discussion.

\medskip
\textbf{Author contributions}


A.~R.: conceptualization, formal analysis, investigation, methodology, project administration, software, validation, visualization, writing~-- original draft, writing~-- review~\& editing

M.~M.: investigation, validation, writing~-- review~\& editing

E.~C.: funding acquisition, writing~-- review~\& editing

Y.~B.: validation, writing~-- review~\& editing

J.~F.: conceptualization, formal analysis, funding acquisition, investigation, methodology, software, supervision, validation, writing~-- original draft, writing~-- review~\& editing

\clearpage

\clearpage

\clearpage
\printbibliography

\appendix
\renewcommand{\thesection}{Appendix \Alph{section}}
\renewcommand{\thesubsection}{\Alph{section}.\arabic{subsection}}
\counterwithin{equation}{section}
\renewcommand{\theequation}{\Alph{section}.\arabic{equation}}

\newpage
\section{\label{app:integer-partition}Integer partition}
The integer partition of the integer $n$ is a way of writing $n$ as a sum of positive integers. The integer partition function $p(n)$ is a count of the number of such ways \cite{euler1753partitione,Rademacher1938}. For example, the integer $p(4) = 5$ as the integer $4$ can be written in five different ways

\begin{align}
    &4,\nonumber\\
    &3 + 1,\nonumber\\
    &2 + 2,\nonumber\\
    &2 + 1 + 2,\nonumber\\
    &1 + 1 + 1 + 1.
    \label{eq:app:integer-partitions-of-4}
\end{align}

From this example one can see that the integer partition is a natural object to classify the terms of the Hamiltonian~\EqAllMightyHamiltonian by the distribution of the spin operators among magnetic sites. For instance, the partition $3+1$ indicates that three out of four spin operators belong to the same magnetic site, while the fourth one belong to the different magnetic site. Thus, the \textit{equivalence class} of interaction parameters is defined as a subset of Hamiltonian's terms within which the amount of spin operators and the associated integer partition are the same.

The terms with single spin operator all belong to the same equivalence class, since there is only one integer partition of $n=1$. 

The terms with two spin operators separate into two equivalence classes: first is associated with the integer partition $2$ (both spin operators belong to the same magnetic site); second is associated with the integer partition $1 + 1$ (the two spin operators are associated with the two distinct magnetic sites). 

The terms with three spin operators split into three equivalence classes ($p(3) = 3$): all are located at the same magnetic site (integer partition $3$); two spins are located on the same site, the third one is located on a different site (integer partition $2 + 1$); all three spins are located at distinct magnetic sites (integer partition $1 + 1 + 1$).

The terms with four spin operators form $p(4) = 5$ equivalence classes in accordance with the five integer partitions listed in Equation~\eqref{eq:app:integer-partitions-of-4}.

\newpage
\section{\label{app:hermiticity}Implications of hermiticity}

The Hamiltonian in Equation~\EqAllMightyHamiltonian of the main text is Hermitian only if the interaction parameters fulfill certain properties. Here we list these properties for the terms with at most four spin components ($n \le 4$). The conditions listed in this section apply equally to both: the initial parameters $J^{i_1,\dots,i_n}_{\nu_2,\dots,\nu_n;\alpha_1,\dots,\alpha_n}$ and the renormalized parameters $\tilde{J}^{i_1,\dots,i_n}_{\nu_2,\dots,\nu_n;\alpha_1,\dots,\alpha_n}$.

The spin Hamiltonian in Equation~\EqAllMightyHamiltonian of the main text can be written as
\begin{equation}
    \mathcal{\hat{H}}
    =
    \sum_{n=1}^{\infty}
    C_n
    \mathcal{\hat{H}}_n,
    \label{eq:app:hermiticity:hamiltonian-as-sum-of-each-n}
\end{equation}
where
\begin{equation}
    \mathcal{\hat{H}}_n
    =
    \sum_{\substack{\mu, \nu_2,\dots,\nu_n,\\ \alpha_1,\dots,\alpha_n,\\ i_1,\dots,i_n}}
    \Biggl(
        J^{i_1,\dots,i_n}_{\nu_2,\dots,\nu_n;\alpha_1,\dots,\alpha_n}
        \prod_{m=1}^{m=n} \hat{S}^{i_m}_{\mu+\nu_m,\alpha_m}
    \Biggr).
    \label{eq:app:hermiticity:hamiltonian_n}
\end{equation}

Since the convention constants $C_n$ are real and each $\mathcal{\hat{H}}_n$ contains exactly $n$ spin operators, the hermiticity of the full Hamiltonian is equivalent to the set of conditions
\begin{equation}
    \mathcal{\hat{H}}_n
    =
    \mathcal{\hat{H}}_n^{\dagger},
    \qquad \forall n.
    \label{eq:app:hermiticity:hermiticity_condition_n}
\end{equation}

Operators $\hat{S}^{x}_{\mu+\nu_m,\alpha_m}$, $\hat{S}^{y}_{\mu+\nu_m,\alpha_m}$, and $\hat{S}^{z}_{\mu+\nu_m,\alpha_m}$ are all Hermitian and interaction parameters are real-valued. Therefore, 

\begin{equation}
    \mathcal{\hat{H}}^{\dagger}_n
    =
    \sum_{\substack{\mu, \nu_2,\dots,\nu_n,\\ \alpha_1,\dots,\alpha_n,\\ i_1,\dots,i_n}}
    \Biggl(
        J^{i_1,\dots,i_n}_{\nu_2,\dots,\nu_n;\alpha_1,\dots,\alpha_n}
        \prod_{m=n}^{m=1} \hat{S}^{i_m}_{\mu+\nu_m,\alpha_m}
    \Biggr).
    \label{eq:app:hermiticity:hamiltonian_n_dagger}
\end{equation}

The properties of the interaction parameters invoked by the hermiticity of the Hamiltonian are obtained by identifying the same combination of spin operators on the left-hand and the right-hand sides of Equation~\eqref{eq:app:hermiticity:hermiticity_condition_n} and requiring the parameters (or sum of parameters) beside them to be equal. We note that the sum over the sets of magnetic sites (i.~e. over $\mu, \nu_2,\dots,\nu_n, \alpha_1,\dots,\alpha_n$) can be separated into two major groups: (i) all magnetic sites are different and (ii) some of the magnetic sites are the same (for example, $\bm{r}_{\mu,\alpha_1} = \bm{r}_{\mu+\nu_m, \alpha_m}$). The spin operators belong to different sites (i.~e. they commute pairwise) for every term from the group (i). Therefore, for the group (i)
\begin{equation}
    \prod_{m=n}^{m=1} \hat{S}^{i_m}_{\mu+\nu_m,\alpha_m}
    =
    \prod_{m=1}^{m=n} \hat{S}^{i_m}_{\mu+\nu_m,\alpha_m}
    \label{eq:app:hermiticity:group-one-product-equivalence}
\end{equation}
and hermiticity of the Hamiltonian poses no properties of the interaction parameters from the group (i).

\subsection{Single spin (n = 1)}

Both sides of Equation~\eqref{eq:app:hermiticity:hermiticity_condition_n} are equivalent. 
This case is trivial as the hermiticity places no properties of the parameters $J_{\alpha_1}^{i_1}$.

\subsection{Two spins (n = 2)}

There is  one type of terms in the group (ii), thus, hermiticity condition $\mathcal{\hat{H}}_2 = \mathcal{\hat{H}}_2^{\dagger}$ simplifies to
\begin{equation}
    \sum_{\mu, \alpha_1,i_1,i_2}
    \Biggl(
        J^{i_1,i_2}_{0;\alpha_1,\alpha_1}
        \hat{S}^{i_1}_{\mu,\alpha_1}
        \hat{S}^{i_2}_{\mu,\alpha_1}
    \Biggr)
    =
    \sum_{\mu, \alpha_1,i_1,i_2}
    \Biggl(
        J^{i_1,i_2}_{0;\alpha_1,\alpha_1}
        \hat{S}^{i_2}_{\mu,\alpha_1}
        \hat{S}^{i_1}_{\mu,\alpha_1}
    \Biggr),
\end{equation}
which implies a single property of the interaction parameters
\begin{equation}
    J^{i_1,i_2}_{0;\alpha_1,\alpha_1} = J^{i_2,i_1}_{0;\alpha_1,\alpha_1}.
    \label{eq:app:hermiticity:parameters-condition_2}
\end{equation}

\subsection{Three spins (n = 3)}

There are two types of terms in the group (ii)

\subsubsection*{Integer partition 3}

All three sites are the same (i.~e. $\bm{r}_{\mu,\alpha_1} = \bm{r}_{\mu+\nu_2,\alpha_2} = \bm{r}_{\mu+\nu_3,\alpha_3}$)

\begin{eqnarray}
    J_{0,0;\alpha_1,\alpha_1,\alpha_1}^{i_1,i_2,i_3}
    =
    J_{0,0;\alpha_1,\alpha_1,\alpha_1}^{i_3,i_2,i_1}
    \label{eq:app:hermiticity:parameters-condition_3_three-same}
\end{eqnarray}

\subsubsection*{Integer partition 2 + 1}

Two sites out of three are the same, but the third one is different.

\begin{eqnarray}
    \Biggl(
        J_{0,\nu_2;\alpha_1,\alpha_1,\alpha_2}^{i_1,i_2,i_3}
        &+&
        J_{\nu_2, 0;\alpha_1,\alpha_2,\alpha_1}^{i_1,i_3,i_2}
        +
        J_{-\nu_2,-\nu_2;\alpha_2,\alpha_1,\alpha_1}^{i_3,i_1,i_2}
    \Biggr)
    =\nonumber\\=
    \Biggl(
        J_{0,\nu_2;\alpha_1,\alpha_1,\alpha_2}^{i_2,i_1,i_3}
        &+&
        J_{\nu_2, 0;\alpha_1,\alpha_2,\alpha_1}^{i_2,i_3,i_1}
        +
        J_{-\nu_2,-\nu_2;\alpha_2,\alpha_1,\alpha_1}^{i_3,i_2,i_1}
    \Biggr),
    \label{eq:app:hermiticity:parameters-condition_3_two-same}
\end{eqnarray}
where Equation~\eqref{eq:app:hermiticity:parameters-condition_3_two-same} only applied when $\bm{r}_{\mu,\alpha_1} \ne \bm{r}_{\mu+\nu_2,\alpha_2}$.

\subsection{Four spins (n = 4)}

There are four types of terms in the group (ii).

\subsubsection*{Integer partition 4}

All four sites are the same (i.~e. $\bm{r}_{\mu,\alpha_1} = \bm{r}_{\mu+\nu_2,\alpha_2} = \bm{r}_{\mu+\nu_3,\alpha_3} = \bm{r}_{\mu+\nu_4,\alpha_4}$)

\begin{equation}
    J_{0,0,0;\alpha_1,\alpha_1,\alpha_1,\alpha_1}^{i_1,i_2,i_3,i_4}
    =
    J_{0,0,0;\alpha_1,\alpha_1,\alpha_1,\alpha_1}^{i_4,i_3,i_2,i_1}.
    \label{eq:app:hermiticity:parameters-condition_4_all-same}
\end{equation}
    
\subsubsection*{Integer partition 3 + 1}

Three sites out of four are the same, but the third one is different.

\begin{eqnarray}
    \Biggl(
        J_{0,0,\nu_2;\alpha_1,\alpha_1,\alpha_1,\alpha_2}^{i_1,i_2,i_3,i_4}
        &+&
        J_{0,\nu_2,0;\alpha_1,\alpha_1,\alpha_2,\alpha_1}^{i_1,i_2,i_4,i_3}
        +\nonumber\\&&+
        J_{\nu_2,0,0;\alpha_1,\alpha_2,\alpha_1,\alpha_1}^{i_1,i_4,i_2,i_3}
        +
        J_{-\nu_2,-\nu_2,-\nu_2;\alpha_2,\alpha_1,\alpha_1,\alpha_1}^{i_4,i_1,i_2,i_3}
    \Biggr)
    =\nonumber\\=
    \Biggl(
        J_{0,0,\nu_2;\alpha_1,\alpha_1,\alpha_1,\alpha_2}^{i_3,i_2,i_1,i_4}
        &+&
        J_{0,\nu_2,0;\alpha_1,\alpha_1,\alpha_2,\alpha_1}^{i_3,i_2,i_4,i_1}
        +\nonumber\\&&+
        J_{\nu_2,0,0;\alpha_1,\alpha_2,\alpha_1,\alpha_1}^{i_3,i_4,i_2,i_1}
        +
        J_{-\nu_2,-\nu_2,-\nu_2;\alpha_2,\alpha_1,\alpha_1,\alpha_1}^{i_4,i_3,i_2,i_1}
    \Biggr),
    \label{eq:app:hermiticity:parameters-condition_4_three-same}
\end{eqnarray}

\noindent where Equation~\eqref{eq:app:hermiticity:parameters-condition_4_three-same} only applied when $\bm{r}_{\mu,\alpha_1} \ne \bm{r}_{\mu+\nu_2,\alpha_2}$.

\subsubsection*{Integer partition 2 + 2}

Two sites out of four are the same, the remaining two sites are the same as well, but different from the first two.

\begin{eqnarray}
    \Biggl(
        J_{0,\nu_2,\nu_2;\alpha_1,\alpha_1,\alpha_2,\alpha_2}^{i_1,i_2,i_3,i_4}
        &+&
        J_{\nu_2,\nu_2,0;\alpha_1,\alpha_2,\alpha_2,\alpha_1}^{i_1,i_3,i_4,i_2}
        +\nonumber\\&+&
        J_{\nu_2,0,\nu_2;\alpha_1,\alpha_2,\alpha_1,\alpha_2}^{i_1,i_3,i_2,i_4}
        +
        J_{-\nu_2,-\nu_2,0;\alpha_2,\alpha_1,\alpha_1,\alpha_2}^{i_3,i_1,i_2,i_4}
        +\nonumber\\&+&
        J_{-\nu_2,0,-\nu_2;\alpha_2,\alpha_1,\alpha_2,\alpha_1}^{i_3,i_1,i_4,i_2}
        +
        J_{0,-\nu_2,-\nu_2;\alpha_2,\alpha_2,\alpha_1,\alpha_1}^{i_3,i_4,i_1,i_2}
    \Biggr)
    =\nonumber\\=
    \Biggl(
        J_{0,\nu_2,\nu_2;\alpha_1,\alpha_1,\alpha_2,\alpha_2}^{i_2,i_1,i_4,i_3}
        &+&
        J_{\nu_2,\nu_2,0;\alpha_1,\alpha_2,\alpha_2,\alpha_1}^{i_2,i_4,i_3,i_1}
        +\nonumber\\&+&
        J_{\nu_2,0,\nu_2;\alpha_1,\alpha_2,\alpha_1,\alpha_2}^{i_2,i_4,i_1,i_3}
        +
        J_{-\nu_2,-\nu_2,0;\alpha_2,\alpha_1,\alpha_1,\alpha_2}^{i_4,i_2,i_1,i_3}
        +\nonumber\\&+&
        J_{-\nu_2,0,-\nu_2;\alpha_2,\alpha_1,\alpha_2,\alpha_1}^{i_4,i_2,i_3,i_1}
        +
        J_{0,-\nu_2,-\nu_2;\alpha_2,\alpha_2,\alpha_1,\alpha_1}^{i_4,i_3,i_2,i_1}
    \Biggr),
    \label{eq:app:hermiticity:parameters-condition_4_two-same_two-same}
\end{eqnarray}

\noindent where Equation~\eqref{eq:app:hermiticity:parameters-condition_4_two-same_two-same} only applied when $\bm{r}_{\mu,\alpha_1} \ne \bm{r}_{\mu+\nu_2,\alpha_2}$.

\subsubsection*{Integer partition 2 + 1 + 1}

Two sites out of four are the same, the remaining two sites are different from each other and from the first two.

\begin{eqnarray}
    \Biggl(
        J_{0,\nu_2,\nu_3;\alpha_1,\alpha_1,\alpha_2,\alpha_3}^{i_1,i_2,i_3,i_4}
        &+&
        J_{0,\nu_3,\nu_2;\alpha_1,\alpha_1,\alpha_3,\alpha_2}^{i_1,i_2,i_4,i_3}
        +\nonumber\\&+&
        J_{\nu_2,0,\nu_3;\alpha_1,\alpha_2,\alpha_1,\alpha_3}^{i_1,i_3,i_2,i_4}
        +
        J_{\nu_3,0,\nu_2;\alpha_1,\alpha_3,\alpha_1,\alpha_2}^{i_1,i_4,i_2,i_3}
        +\nonumber\\&+&
        J_{\nu_2,\nu_3,0;\alpha_1,\alpha_2,\alpha_3,\alpha_1}^{i_1,i_3,i_4,i_2}
        +
        J_{\nu_3,\nu_2,0;\alpha_1,\alpha_3,\alpha_2,\alpha_1}^{i_1,i_4,i_3,i_2}
        +\nonumber\\&+&
        J_{-\nu_2,-\nu_2,\nu_3-\nu_2;\alpha_2,\alpha_1,\alpha_1,\alpha_3}^{i_3,i_1,i_2,i_4}
        +
        J_{-\nu_3,-\nu_3,\nu_2-\nu_3;\alpha_3,\alpha_1,\alpha_1,\alpha_2}^{i_4,i_1,i_2,i_3}
        +\nonumber\\&+&
        J_{-\nu_2,\nu_3-\nu_2,-\nu_2;\alpha_2,\alpha_1,\alpha_3,\alpha_1}^{i_3,i_1,i_4,i_2}
        +
        J_{-\nu_3,\nu_2-\nu_3,-\nu_3;\alpha_3,\alpha_1,\alpha_2,\alpha_1}^{i_4,i_1,i_3,i_2}
        +\nonumber\\&+&
        J_{\nu_3-\nu_2,-\nu_2,-\nu_2;\alpha_2,\alpha_3,\alpha_1,\alpha_1}^{i_3,i_4,i_1,i_2}
        +
        J_{\nu_2-\nu_3,-\nu_3,-\nu_3;\alpha_3,\alpha_2,\alpha_1,\alpha_1}^{i_4,i_3,i_1,i_2}
    \Biggr)
    =\nonumber\\=
    \Biggl(
        J_{0,\nu_2,\nu_3;\alpha_1,\alpha_1,\alpha_2,\alpha_3}^{i_2,i_1,i_3,i_4}
        &+&
        J_{0,\nu_3,\nu_2;\alpha_1,\alpha_1,\alpha_3,\alpha_2}^{i_2,i_1,i_4,i_3}
        +\nonumber\\&+&
        J_{\nu_2,0,\nu_3;\alpha_1,\alpha_2,\alpha_1,\alpha_3}^{i_2,i_3,i_1,i_4}
        +
        J_{\nu_3,0,\nu_2;\alpha_1,\alpha_3,\alpha_1,\alpha_2}^{i_2,i_4,i_1,i_3}
        +\nonumber\\&+&
        J_{\nu_2,\nu_3,0;\alpha_1,\alpha_2,\alpha_3,\alpha_1}^{i_2,i_3,i_4,i_1}
        +
        J_{\nu_3,\nu_2,0;\alpha_1,\alpha_3,\alpha_2,\alpha_1}^{i_2,i_4,i_3,i_1}
        +\nonumber\\&+&
        J_{-\nu_2,-\nu_2,\nu_3-\nu_2;\alpha_2,\alpha_1,\alpha_1,\alpha_3}^{i_3,i_2,i_1,i_4}
        +
        J_{-\nu_3,-\nu_3,\nu_2-\nu_3;\alpha_3,\alpha_1,\alpha_1,\alpha_2}^{i_4,i_2,i_1,i_3}
        +\nonumber\\&+&
        J_{-\nu_2,\nu_3-\nu_2,-\nu_2;\alpha_2,\alpha_1,\alpha_3,\alpha_1}^{i_3,i_2,i_4,i_1}
        +
        J_{-\nu_3,\nu_2-\nu_3,-\nu_3;\alpha_3,\alpha_1,\alpha_2,\alpha_1}^{i_4,i_2,i_3,i_1}
        +\nonumber\\&+&
        J_{\nu_3-\nu_2,-\nu_2,-\nu_2;\alpha_2,\alpha_3,\alpha_1,\alpha_1}^{i_3,i_4,i_2,i_1}
        +
        J_{\nu_2-\nu_3,-\nu_3,-\nu_3;\alpha_3,\alpha_2,\alpha_1,\alpha_1}^{i_4,i_3,i_2,i_1}
    \Biggr),
    \label{eq:app:hermiticity:parameters-condition_4_two-same_two-different}
\end{eqnarray}

\noindent where Equation~\eqref{eq:app:hermiticity:parameters-condition_4_two-same_two-different} only applied when $\bm{r}_{\mu,\alpha_1}$, 
$\bm{r}_{\mu+\nu_2,\alpha_2}$, $\bm{r}_{\mu+\nu_3,\alpha_3}$ are all different.

\subsection{Arbitrary n}

Note that the amount of types in the group (ii) for any $n$ is equal to $p(n) -1$, where $p(n)$ is an integer partition function. The analytical form of the integer partition function is not know which directly translates into the fact that we can not write a general expression for the Hermiticity condition, or equivalent sets (\ref{app:gauge}), or $E^{corr}$ and eigenstate conditions (\ref{app:vacuum-state}), or renormalized parameters (Equation~\EqRenormalizedParameters of the main text and~\ref{app:renormalization}) for an arbitrary $n$ as these quantities have different forms for each equivalent class. Nevertheless, an algorithm for writing down all that quantities is the same as the one we use for $n \le 4$.

\newpage
\section{\label{app:gauge}Gauge invariance}
Before explaining the gauge invariance for arbitrary $n$ we give an example for the terms with $n=2$ and integer partition $1+1$. The choice between symmetric gauge and minimal gauge for these terms is known in the literature and the choice between ``double counting'' and ``no double counting''.

\subsection{An example}

Consider two spins located at different positions in real space, $\bm{\hat{S}}_1$ and $\bm{\hat{S}}_2$, that are coupled via the $3\times3$ matrices of the interaction parameters, $\bm{J}_{12}$ and $\bm{J}_{21}$, and assume that the interaction parameters between these two spins have the form

\begin{eqnarray}
    \bm{J}_{12}
    =
    \begin{pmatrix}
        0 & a & 0 \\
        b & 0 & 0 \\
        0 & 0 & 0 \\
    \end{pmatrix};
    \qquad
    \bm{J}_{21}
    =
    \begin{pmatrix}
        0 & c & 0 \\
        d & 0 & 0 \\
        0 & 0 & 0 \\
    \end{pmatrix},
\end{eqnarray}
where no restrictions are placed on the real numbers $a,b,c,d$. When the sum over magnetic sites in the $n=2$ term of the Hamiltonian in Equation~\EqAllMightyHamiltonian of the main text is expanded, the bilinear interaction between these two spins becomes (recall that spin operators of different sites commute: $[\hat{S}_1^i, \hat{S}_2^j] = 0$)

\begin{eqnarray}
    \mathcal{\hat{H}}
    &=&
    \ldots
    +
    a \hat{S}_1^x \hat{S}_2^y
    +
    b \hat{S}_1^y \hat{S}_2^x
    +
    c \hat{S}_2^x \hat{S}_1^y
    +
    d \hat{S}_2^y \hat{S}_1^x
    +
    \ldots
    =\nonumber\\&&=
    \ldots
    +
    (a+d) \hat{S}_1^x \hat{S}_2^y
    +
    (b+c) \hat{S}_1^y \hat{S}_2^x
    +
    \ldots
\end{eqnarray}
where only the sum $a+d$ (or $b+c$) defines the behavior of the Hamiltonian. Therefore, the sum can be arbitrarily distributed between $J_{12}^{xy}$ and $J_{21}^{yx}$ (or $J_{12}^{yx}$ and $J_{21}^{xy}$) without modifying the Hamiltonian. There are two popular choices of such distribution that dominate the literature
\begin{equation}
    \bm{J}_{12}
    =
    \bm{J}_{21}^T
    =
    \begin{pmatrix}
        0 & \dfrac{a+d}{2} & 0 \\
        \dfrac{b+c}{2} & 0 & 0 \\
        0 & 0 & 0 \\
    \end{pmatrix}
\end{equation}

\noindent or

\begin{equation}
    \bm{J}_{12}
    =
    \begin{pmatrix}
        0 & a+d & 0 \\
        b+c & 0 & 0 \\
        0 & 0 & 0 \\
    \end{pmatrix},
    \qquad
    \bm{J}_{21}
    =
    \bm{0}.
\end{equation}

In general, the term gauge invariance or gauge freedom implies that there is some redundancy in the mathematical description of the model that does not affect physics of the modeled system. That meaning captures the essence of the example from above and generalizes for arbitrary $n$.

\subsection{Generalization for an arbitrary n}

First, given a combination of spin operators we define an invariant set of interaction parameters as the set of all tensor components of interaction parameters whose associated combination of the spin operators is identical to the given one up to permutation of commuting spin operators.

Then the gauge transformation is defined as the redistribution of the tensor components within each invariant set, while leaving the sum over each invariant set unchanged, i.e.

\begin{equation}
  J_a \to J_a + \delta_a, \qquad \sum_a \delta_a = 0,
\end{equation}

\noindent where compound index $a$ runs over elements of the invariant set.

The PLSM Hamiltonian is gauge invariant by definition. There are two gauge choices widely used in the literature. First, \textit{symmetric gauge}, where for any invariant set $\mathcal{A}$ its elements are equal to one another, i.e.

\begin{equation}
    J_a = J_b, \qquad \forall a,b \in \mathcal{A},\qquad \forall \mathcal{A}.
\end{equation}

\noindent Second, \textit{minimal gauge}, where for any invariant set $\mathcal{A}$ there is only one element with non-zero value, i.e.

\begin{equation}
    \exists !\quad a_0 \in \mathcal{A}: 
    J_{a_0} \ne 0 \land J_a = 0\quad \forall a \ne a_0,
    \qquad
    \forall \mathcal{A}
\end{equation}

The method of listing the type of invariant set for any chosen $n$ and integer partition of $n$ based on two observations: the spin operators associated with the different sites can be swapped; the site indices and indices of Cartesian components of the pairs of spin operators associated with the same site can be renamed.

The equivalence class for any $n$ and all spin operators belonging to the same site can only host a trivial invariant sets with a single element. Thus, for any $n$ there are $3^n$ invariant sets of the form

\begin{equation}
    \{J_{0, \dots,0;\alpha_1,\dots,\alpha_n}^{i_1,\dots,i_n}\}
\end{equation}

For $n = 1$ there is only one equivalence class, thus, only three sets of the form from above are possible. For the $2 \le n \le 4$ we list the invariant sets below.

\subsection{One spin (n=1)}

This case is trivial as each equivalent set contains a single parameter. Equal distribution does not imply any properties of the parameters $J_{\alpha}^{i}$.

\subsection{Two spins (n = 2)}

A single equivalence class remains to be described, where two sites are different (i.~e. $\bm{r}_{\mu,\alpha_1} \ne \bm{r}_{\mu+\nu_2, \alpha_2}$).

\begin{equation}
    \Bigl\{
        J^{i_1,i_2}_{\nu_2;\alpha_1,\alpha_2},
        J^{i_2,i_1}_{-\nu_2;\alpha_2,\alpha_1}
    \Bigr\}.
    \label{eq:app:equivalence:set_2}
\end{equation}

Note that the invariant sets in Equation~\eqref{eq:app:equivalence:set_2} are only defined when $\bm{r}_{\mu,\alpha_1} \ne \bm{r}_{\mu+\nu_2, \alpha_2}$.

\subsection{Three spins (n = 3)}

\subsubsection*{Integer partition 2 + 1}

Two sites out of three are the same, but the third one is different.

\begin{equation}
    \Bigl\{
        J_{0,\nu_2;\alpha_1,\alpha_1,\alpha_2}^{i_1,i_2,i_3},\quad
        J_{\nu_2,0;\alpha_1,\alpha_2,\alpha_1}^{i_1,i_3,i_2},\quad
        J_{-\nu_2,-\nu_2;\alpha_2,\alpha_1,\alpha_1}^{i_3,i_1,i_2}
    \Bigr\},
    \label{eq:app:equivalence:set_3_two_diff}
\end{equation}
where invariant sets in Equation~\eqref{eq:app:equivalence:set_3_two_diff} are only defined when $\bm{r}_{\mu,\alpha_1} \ne \bm{r}_{\mu+\nu_2,\alpha_2}$. 

\subsubsection*{Integer partition 1 + 1 + 1}

All three sites are different.

\begin{eqnarray}
    &&\Bigl\{
        J_{\nu_2,\nu_3;\alpha_1,\alpha_2,\alpha_3}^{i_1,i_2,i_3},\quad
        J_{\nu_3,\nu_2;\alpha_1,\alpha_3,\alpha_2}^{i_1,i_3,i_2},\quad
        \nonumber\\&&\phantom{\Bigl\{}
        J_{-\nu_2,\nu_3-\nu_2;\alpha_2,\alpha_1,\alpha_3}^{i_2,i_1,i_3},\quad
        J_{\nu_3-\nu_2,-\nu_2;\alpha_2,\alpha_3,\alpha_1}^{i_2,i_3,i_1},\quad
        \nonumber\\&&\phantom{\Bigl\{}
        J_{-\nu_3,\nu_2-\nu_3;\alpha_3,\alpha_1,\alpha_2}^{i_3,i_1,i_2},\quad
        J_{\nu_2-\nu_3,-\nu_3;\alpha_3,\alpha_2,\alpha_1}^{i_3,i_2,i_1}
    \Bigr\},
    \label{eq:app:equivalence:set_3_all_diff}
\end{eqnarray}
where invariant sets in Equation~\eqref{eq:app:equivalence:set_3_all_diff} are only defined when $\bm{r}_{\mu,\alpha_1}$, $\bm{r}_{\mu+\nu_2,\alpha_2}$, and $\bm{r}_{\mu+\nu_3,\alpha_3}$ are all different sites.

\subsection{Four spins (n=4)}
    
\subsubsection*{Integer partition 3 + 1}

Three sites out of four are the same, but the third one is different.

\begin{eqnarray}
    &&\Bigl\{
        J_{0,0,\nu_2;\alpha_1,\alpha_1,\alpha_1,\alpha_2}^{i_1,i_2,i_3,i_4},\quad
        J_{0,\nu_2,0;\alpha_1,\alpha_1,\alpha_2,\alpha_1}^{i_1,i_2,i_4,i_3},\quad
        \nonumber\\&&\phantom{\Bigl\{}
        J_{\nu_2,0,0;\alpha_1,\alpha_2,\alpha_1,\alpha_1}^{i_1,i_4,i_2,i_3},\quad
        J_{-\nu_2,-\nu_2,-\nu_2;\alpha_2,\alpha_1,\alpha_1,\alpha_1}^{i_4,i_1,i_2,i_3}
    \Bigr\},
    \label{eq:app:equivalence:set_4_one_diff}
\end{eqnarray}
where invariant sets in Equation~\eqref{eq:app:equivalence:set_4_two_same_two_same} are only defined when $\bm{r}_{\mu,\alpha_1} \ne   \bm{r}_{\mu+\nu_2,\alpha_2}$. 

\subsubsection*{Integer partition 2 + 2}

Two sites out of four are the same, the remaining two sites are the same as well, but different from the first two.

\begin{eqnarray}
    &&\Bigl\{
        J_{0,\nu_2,\nu_2;\alpha_1,\alpha_1,\alpha_2,\alpha_2}^{i_1,i_2,i_3,i_4},\quad
        J_{\nu_2,0,\nu_2;\alpha_1,\alpha_2,\alpha_1,\alpha_2}^{i_1,i_3,i_2,i_4},\quad
        \nonumber\\&&\phantom{\Bigl\{}
        J_{\nu_2,\nu_2,0;\alpha_1,\alpha_2,\alpha_2,\alpha_1}^{i_1,i_3,i_4,i_2},\quad
        J_{-\nu_2,-\nu_2,0;\alpha_2,\alpha_1,\alpha_1,\alpha_2}^{i_3,i_1,i_2,i_4},\quad
        \nonumber\\&&\phantom{\Bigl\{}
        J_{-\nu_2,0,-\nu_2;\alpha_2,\alpha_1,\alpha_2,\alpha_1}^{i_3,i_1,i_4,i_2},\quad
        J_{0,-\nu_2,-\nu_2;\alpha_2,\alpha_2,\alpha_1,\alpha_1}^{i_3,i_4,i_1,i_2}
    \Bigr\},
    \label{eq:app:equivalence:set_4_two_same_two_same}
\end{eqnarray}
where invariant sets in Equation~\eqref{eq:app:equivalence:set_4_one_diff} are only defined when $\bm{r}_{\mu,\alpha_1} \ne   \bm{r}_{\mu+\nu_2,\alpha_2}$. 

\subsubsection*{Integer partition 2 + 1 + 1}

Two sites out of four are the same, the remaining two sites are different from each other and from the first two.

\begin{eqnarray}
    &&\Bigl\{
        J_{0,\nu_2,\nu_3;\alpha_1,\alpha_1,\alpha_2,\alpha_3}^{i_1,i_2,i_3,i_4},\quad
        J_{0,\nu_3,\nu_2;\alpha_1,\alpha_1,\alpha_3,\alpha_2}^{i_1,i_2,i_4,i_3},\quad
        \nonumber\\&&\phantom{\Bigl\{}
        J_{\nu_2,0,\nu_3;\alpha_1,\alpha_2,\alpha_1,\alpha_3}^{i_1,i_3,i_2,i_4},\quad
        J_{\nu_3,0,\nu_2;\alpha_1,\alpha_3,\alpha_1,\alpha_2}^{i_1,i_4,i_2,i_3},\quad
        \nonumber\\&&\phantom{\Bigl\{}
        J_{\nu_2,\nu_3,0;\alpha_1,\alpha_2,\alpha_3,\alpha_1}^{i_1,i_3,i_4,i_2},\quad
        J_{\nu_3,\nu_2,0;\alpha_1,\alpha_3,\alpha_2,\alpha_1}^{i_1,i_4,i_3,i_2},\quad
        \nonumber\\&&\phantom{\Bigl\{}
        J_{-\nu_2,-\nu_2,\nu_3-\nu_2;\alpha_2,\alpha_1,\alpha_1,\alpha_3}^{i_3,i_1,i_2,i_4},\quad
        J_{-\nu_3,-\nu_3,\nu_2-\nu_3;\alpha_3,\alpha_1,\alpha_1,\alpha_2}^{i_4,i_1,i_2,i_3},\quad
        \nonumber\\&&\phantom{\Bigl\{}
        J_{-\nu_2,\nu_3-\nu_2,-\nu_2;\alpha_2,\alpha_1,\alpha_3,\alpha_1}^{i_3,i_1,i_4,i_2},\quad
        J_{-\nu_3,\nu_2-\nu_3,-\nu_3;\alpha_3,\alpha_1,\alpha_2,\alpha_1}^{i_4,i_1,i_3,i_2},\quad
        \nonumber\\&&\phantom{\Bigl\{}
        J_{\nu_3-\nu_2,-\nu_2,-\nu_2;\alpha_2,\alpha_3,\alpha_1,\alpha_1}^{i_3,i_4,i_1,i_2},\quad
        J_{\nu_2-\nu_3,-\nu_3,-\nu_3;\alpha_3,\alpha_2,\alpha_1,\alpha_1}^{i_4,i_3,i_1,i_2}
    \Bigr\},
    \label{eq:app:equivalence:set_4_two_diff_two_same}
\end{eqnarray}
where einvariant sets in Equation~\eqref{eq:app:equivalence:set_4_two_diff_two_same} are only defined when $\bm{r}_{\mu,\alpha_1}$, $\bm{r}_{\mu+\nu_2,\alpha_2}$, and $\bm{r}_{\mu+\nu_3,\alpha_3}$ are all different sites.

\subsubsection*{Integer partition 1 + 1 + 1 + 1}

All four sites are different.

\begin{eqnarray}
    &&\Bigl\{
        J_{\nu_2,\nu_3,\nu_4;\alpha_1,\alpha_2,\alpha_3,\alpha_4}^{i_1,i_2,i_3,i_4},\quad
        J_{\nu_2,\nu_4,\nu_3;\alpha_1,\alpha_2,\alpha_4,\alpha_3}^{i_1,i_2,i_4,i_3},\quad
        \nonumber\\&&\phantom{\Bigl\{}
        J_{\nu_3,\nu_2,\nu_4;\alpha_1,\alpha_3,\alpha_2,\alpha_4}^{i_1,i_3,i_2,i_4},\quad
        J_{\nu_3,\nu_4,\nu_2;\alpha_1,\alpha_3,\alpha_4,\alpha_2}^{i_1,i_3,i_4,i_2},\quad
        \nonumber\\&&\phantom{\Bigl\{}
        J_{\nu_4,\nu_2,\nu_3;\alpha_1,\alpha_4,\alpha_2,\alpha_3}^{i_1,i_4,i_2,i_3},\quad
        J_{\nu_4,\nu_3,\nu_2;\alpha_1,\alpha_4,\alpha_3,\alpha_2}^{i_1,i_4,i_3,i_2},\quad
        \nonumber\\&&\phantom{\Bigl\{}
        J_{-\nu_2,\nu_3-\nu_2,\nu_4-\nu_2;\alpha_2,\alpha_1,\alpha_3,\alpha_4}^{i_2,i_1,i_3,i_4},\quad
        J_{-\nu_2,\nu_4-\nu_2,\nu_3-\nu_2;\alpha_2,\alpha_1,\alpha_4,\alpha_3}^{i_2,i_1,i_4,i_3},\quad
        \nonumber\\&&\phantom{\Bigl\{}
        J_{\nu_3-\nu_2,-\nu_2,\nu_4-\nu_2;\alpha_2,\alpha_3,\alpha_1,\alpha_4}^{i_2,i_3,i_1,i_4},\quad
        J_{\nu_3-\nu_2,\nu_4-\nu_2,-\nu_2;\alpha_2,\alpha_3,\alpha_4,\alpha_1}^{i_2,i_3,i_4,i_1},\quad
        \nonumber\\&&\phantom{\Bigl\{}
        J_{\nu_4-\nu_2,-\nu_2,\nu_3-\nu_2;\alpha_2,\alpha_4,\alpha_1,\alpha_3}^{i_2,i_4,i_1,i_3},\quad
        J_{\nu_4-\nu_2,\nu_3-\nu_2,-\nu_2;\alpha_2,\alpha_4,\alpha_3,\alpha_1}^{i_2,i_4,i_3,i_1},\quad
        \nonumber\\&&\phantom{\Bigl\{}
        J_{-\nu_3,\nu_2-\nu_3,\nu_4-\nu_3;\alpha_3,\alpha_1,\alpha_2,\alpha_4}^{i_3,i_1,i_2,i_4},\quad
        J_{-\nu_3,\nu_4-\nu_3,\nu_2-\nu_3;\alpha_3,\alpha_1,\alpha_4,\alpha_2}^{i_3,i_1,i_4,i_2},\quad
        \nonumber\\&&\phantom{\Bigl\{}
        J_{\nu_2-\nu_3,-\nu_3,\nu_4-\nu_3;\alpha_3,\alpha_2,\alpha_1,\alpha_4}^{i_3,i_2,i_1,i_4},\quad
        J_{\nu_2-\nu_3,\nu_4-\nu_3,-\nu_3;\alpha_3,\alpha_2,\alpha_4,\alpha_1}^{i_3,i_2,i_4,i_1},\quad
        \nonumber\\&&\phantom{\Bigl\{}
        J_{\nu_4-\nu_3,-\nu_3,\nu_2-\nu_3;\alpha_3,\alpha_4,\alpha_1,\alpha_2}^{i_3,i_4,i_1,i_2},\quad
        J_{\nu_4-\nu_3,\nu_2-\nu_3,-\nu_3;\alpha_3,\alpha_4,\alpha_2,\alpha_1}^{i_3,i_4,i_2,i_1},\quad
        \nonumber\\&&\phantom{\Bigl\{}
        J_{-\nu_4,\nu_2-\nu_4,\nu_3-\nu_4;\alpha_4,\alpha_1,\alpha_2,\alpha_3}^{i_4,i_1,i_2,i_3},\quad
        J_{-\nu_4,\nu_3-\nu_4,\nu_2-\nu_4;\alpha_4,\alpha_1,\alpha_3,\alpha_2}^{i_4,i_1,i_3,i_2},\quad
        \nonumber\\&&\phantom{\Bigl\{}
        J_{\nu_2-\nu_4,-\nu_4,\nu_3-\nu_4;\alpha_4,\alpha_2,\alpha_1,\alpha_3}^{i_4,i_2,i_1,i_3},\quad
        J_{\nu_2-\nu_4,\nu_3-\nu_4,-\nu_4;\alpha_4,\alpha_2,\alpha_3,\alpha_1}^{i_4,i_2,i_3,i_1},\quad
        \nonumber\\&&\phantom{\Bigl\{}
        J_{\nu_3-\nu_4,-\nu_4,\nu_2-\nu_4;\alpha_4,\alpha_3,\alpha_1,\alpha_2}^{i_4,i_3,i_1,i_2},\quad
        J_{\nu_3-\nu_4,\nu_2-\nu_4,-\nu_4;\alpha_4,\alpha_3,\alpha_2,\alpha_1}^{i_4,i_3,i_2,i_1}
    \Bigr\},
    \label{eq:app:equivalence:set_4_all_diff}
\end{eqnarray}
where invariant sets in Equation~\eqref{eq:app:equivalence:set_4_all_diff} are only defined when $\bm{r}_{\mu,\alpha_1}$, $\bm{r}_{\mu+\nu_2,\alpha_2}$, $\bm{r}_{\mu+\nu_3,\alpha_3}$, and $\bm{r}_{\mu+\nu_4,\alpha_4}$ are all different sites.

\subsection{Symmetric gauge with hermiticity}

In the special case of $n=2$ the hermiticity condition (Equation~\eqref{eq:app:hermiticity:parameters-condition_2}) and the symmetric gauge combine into a single property

\begin{equation}
    J^{i_1,i_2}_{\nu_2;\alpha_1,\alpha_2}
    =
    J^{i_2,i_1}_{-\nu_2;\alpha_2,\alpha_1}.
    \label{eq:app:equivalence:2_combined_with_hermiticity}
\end{equation}

However, for $n \ge 3$ hermiticity and the gauge choice can not be combined as nicely. Therefore, one shall remember that these two properties not only imply different properties of interaction parameters, but, perhaps more importantly, have distinct nature. Hermiticity is a physical property. Gauge is a convention choice.

\newpage
\section{\label{app:vacuum-state}Quantum energy correction and eigenstate condition}

Expectation value of the Hamiltonian in Equation~\EqAllMightyHamiltonian of the main text in the state $\ket{0}\equiv\ket{S^{z_{\alpha}}_{\mu, \alpha} = S_{\alpha}, \forall\mu, \forall\alpha}$ as well as the condition for $\ket{0}$ to be an eigenstate of Hamiltonian are evaluated directly by using the renormalized form of the Hamiltonian in Equation~\EqRenormalizedAllMightyHamiltonian of the main text, the exact representation of spin deviation operator
\begin{equation}
    \bm{\hat{\Delta}}_{\mu, \alpha}
    =
    \dfrac{\overline{\bm{p}_{\alpha}}}{2}
    \hat{S}^+_{\mu, \alpha}
    +
    \dfrac{\bm{p}_{\alpha}}{2}
    \hat{S}^-_{\mu, \alpha}
    +
    \Bigl(
        \hat{S}_{\mu,\alpha}^{z_{\alpha}}
        -
        S_{\alpha}
    \Bigr)
    \bm{z}_{\alpha},
    \label{eq:app:vacuum-state:exact_repr_of_deviation_operator}
\end{equation}
and the action of spin operators on the eigenstates of $\hat{S}_{\mu,\alpha}^{z_{\alpha}}$
\begin{eqnarray}
    \hat{S}_{\mu,\alpha}^{z_{\alpha}} \ket{S_{\mu,\alpha}^{z_{\alpha}}= S_{\alpha}^{z_{\alpha}}}
    &=&
    S_{\alpha}^{z_{\alpha}}\ket{S_{\mu,\alpha}^{z_{\alpha}}=S_{\alpha}^{z_{\alpha}}}
    \\
    \hat{S}_{\mu,\alpha}^{\pm} \ket{S_{\mu,\alpha}^{z_{\alpha}}=S_{\alpha}^{z_{\alpha}}}
    &=&
    \sqrt{
        S_{\alpha}(S_{\alpha} + 1)
        -
        S_{\alpha}^{z_{\alpha}}(S_{\alpha}^{z_{\alpha}} \pm 1)
    }\ket{S_{\mu,\alpha}^{z_{\alpha}} = S_{\alpha}^{z_{\alpha}} \pm 1},
    \label{eq:app:vacuum-state:action_spin_operators}
\end{eqnarray}
where $S_{\alpha}^{z_{\alpha}} = -S_{\alpha}, \ldots, S_{\alpha}$ are (half-)integer numbers.

As we mention in~\ref{app:hermiticity}, the calculation of $E^{corr}$ and of the eigenstate condition is deeply intertwined with the concept of integer partition. Thus, we can not provide the analytical form of either for an arbitrary $n$. Instead we list both of them for $n \le 4$.

\subsection{Expectation value}

Full quantum correction to the classical part ($E^{(0)}$) of the expectation value $\bra{0}\mathcal{\hat{H}}\ket{0}$ can be expressed as
\begin{equation}
    E^{corr}
    =
    \sum_{n=1}^{\infty}
    E^{corr}_n.
\end{equation}

First four terms of that sum are

\begin{subequations}
\begin{eqnarray}
    E_{n=1}^{corr} &=& 0,
    \label{eq:app:vacuum-state:E_1^corr}
    \\
    E_{n=2}^{corr}
    &=&
    \sum_{\alpha_1,i_1,i_2}
    \dfrac{S_{\alpha_1}}{2}
    \tilde{J}^{i_1,i_2}_{0;\alpha_1,\alpha_1}
    \overline{p_{\alpha_1}^{i_1}}
    p_{\alpha_1}^{i_2},
    \label{eq:app:vacuum-state:E_2^corr}
    \\
    E_{n=3}^{corr}
    &=&
    -
    \sum_{\substack{\alpha_1,i_1,\\i_2,i_3}}
    \dfrac{S_{\alpha_1}}{2}
    \tilde{J}^{i_1,i_2,i_3}_{0,0;\alpha_1,\alpha_1,\alpha_1}
    \overline{p_{\alpha_1}^{i_1}}
    z_{\alpha_1}^{i_2}
    p_{\alpha_1}^{i_3},
    \label{eq:app:vacuum-state:E_3^corr}
    \\
    E_{n=4}^{corr}
    &=&
    \sum_{\substack{\alpha_1,i_1,\\i_2,i_3,i_4}}
    \dfrac{S_{\alpha_1}}{2}
    \Biggl(
        \tilde{J}^{i_1,\dots,i_4}_{0,0,0;\alpha_1,\alpha_1,\alpha_1,\alpha_1}
        \overline{p_{\alpha_1}^{i_1}}
        \Biggl(
            z_{\alpha_1}^{i_2}
            z_{\alpha_1}^{i_3}
            +
            \dfrac{1}{2} S_{\alpha_1}
            p_{\alpha_1}^{i_2}
            \overline{p_{\alpha_1}^{i_3}}
            +
            \left(S_{\alpha_1} - \dfrac{1}{2}\right)
            \overline{p_{\alpha_1}^{i_2}}
            p_{\alpha_1}^{i_3}
        \Biggr)
        p_{\alpha_1}^{i_4}
        +\nonumber\\&&\phantom{\sum_{\substack{\alpha_1,i_1,\\i_2,i_3,i_4}}\dfrac{S_{\alpha_1}}{2}\Biggl(}+
        \sum_{\nu_2, \alpha_2}
        \dfrac{3}{2}
        S_{\alpha_2}
        \tilde{J}^{i_1,\dots,i_4}_{0,\nu_2,\nu_2;\alpha_1,\alpha_1,\alpha_2,\alpha_2}
        \overline{p_{\alpha_1}^{i_1}}
        p_{\alpha_1}^{i_2}
        \overline{p_{\alpha_2}^{i_3}}
        p_{\alpha_2}^{i_4}
        [\bm{r}_{\mu,\alpha_1} \ne \bm{r}_{\mu+\nu_2,\alpha_2}]
    \Biggr),
    \label{eq:app:vacuum-state:E_4^corr}
\end{eqnarray}
\label{eq:app:vacuum-state:E^corr}
\end{subequations}
where $[\bm{r}_{\mu,\alpha_1} \ne \bm{r}_{\mu+\nu_2,\alpha_2}]$ is an Iverson bracket \cite{Iverson1962-us,Knuth1992}.

\begin{equation}
    [\bm{r}_{\mu,\alpha_1} \ne \bm{r}_{\mu+\nu_2,\alpha_2}]
    =
    \left\{
    \begin{matrix}
        1  \text{, if } \bm{r}_{\mu,\alpha_1} \ne \bm{r}_{\mu+\nu_2,\alpha_2}, \\
        0 \text{, if } \bm{r}_{\mu,\alpha_1} = \bm{r}_{\mu+\nu_2,\alpha_2}
    \end{matrix}
    \right..
    \label{eq:app:vacuum-state:iverson_bracket}
\end{equation}

The Equation~\eqref{eq:app:vacuum-state:E^corr} are listed under the assumption of the symmetric gauge of the renormalized parameters as defined in~\ref{app:gauge}.

\subsection{Eigenstate conditions}

Action of the operator $\mathcal{\hat{H}}^{corr}\equiv \mathcal{\hat{H}} - NE^{(0)}$ on $\ket{0}$ produces a linear combination of states from the Hilbert space $\mathcal{L}$ (defined in the main text). We classify these states as follows. For a fixed $n$, two states are said to be of the same type if they belong to the same integer partition of $n$ (for the definition of \text{integer partition} see, for instance, \cite{euler1753partitione,Rademacher1938}). For example, the states $\ket{S_{\mu,\alpha_1}^{z_{\alpha_1}} = S_{\alpha_1} - 2}$, $\ket{S_{\mu,\alpha_1}^{z_{\alpha_1}} = S_{\alpha_1} - 1, S_{\mu+\nu_2,\alpha_2}^{z_{\alpha_2}} = S_{\alpha_2} - 1}$, and $\ket{S_{\mu+\nu_2,\alpha_2}^{z_{\alpha_2}} = S_{\alpha_2} - 2}$ are said to belong to the integer partition $(2+0)$ of the integer $2$. Then, the total amount of \textit{types} of those states for each $n$ is simply given by the integer partition function $p(n)$. Note that the vacuum state formally belongs to the integer partition $(0)$ of the integer $0$. Therefore, the total amount of types of states that can be achieved by the action of $\mathcal{\hat{H}}$ on $\ket{0}$, assuming that the Hamiltonian contains terms with at most $n$ spin operators, can be computed as
\begin{equation}
    N_{types}(n) = \sum_{m=0}^{m=n} p(n).
    \label{eq:app:vacuum-state:N_types_n}
\end{equation}

The values of $p(n)$ and $N_{types}(n)$ for $1 \le n \le 4$ are summarized in the Table~\ref{table:app:vacuum-state:p_N_values}

\begin{table}[h]
\centering
\def\arraystretch{1.5}
\setlength{\tabcolsep}{10pt}
\caption{\label{table:app:vacuum-state:p_N_values} Values of integer partition function $p(n)$ and $N_{types}(n)$ for $1 \le n \le 4$.}
\begin{tabular}{c|cc}
    $n$ & $p(n)$ & $N_{types}(n)$ \\ \hline
    1 & 1 & 2 \\
    2 & 2 & 4 \\
    3 & 3 & 7 \\
    4 & 5 & 12 \\
\end{tabular}
\end{table}

There are $12$ types of states that appear in the expression for $\mathcal{\hat{H}}^{corr}_{n\le 4}\ket{0}$, thus, it can be written as

\begin{eqnarray}
    \mathcal{\hat{H}}^{corr}_{n\le 4}\ket{0}
    &=&
    NE^{corr}_{n\le 4}\ket{0}
    +
    \sum_{u=1}^{u=11}
    E^{u}_{n\le 4}\ket{u}
    \label{eq:app:vacuum-state:H_corr_action}
\end{eqnarray}
where the composite index $u$ hides the summation over all states of the same type $\ket{u}$. The state $\ket{0}$ is an eigenstate of the Hamiltonian in Equation~\EqAllMightyHamiltonian of the main text (assuming $n \le 4$) if $E^{u}_{n \le 4} = 0$ for all $1 \le u \le 11$. The values of $E^{u}_{n \le 4}$ are listed below. We use Iverson brackets \cite{Iverson1962-us,Knuth1992} to indicate conditions on the site's positions. The Equation~\eqref{eq:app:vacuum-state:eigenstate_condition_n_4_u_1}~--~\eqref{eq:app:vacuum-state:eigenstate_condition_n_4_u_11} and Equation~\eqref{eq:app:vacuum-state:eigenstate_condition_n_2_u_123} are listed under the assumption of the symmetric gauge of the renormalized parameters as defined in~\ref{app:gauge}. In the Table~\ref{table:app:vacuum-state:E_n4_u} we specify the connection between values of index $u$ and type of state it represents.

\newpage
{
\newlength\rowgap \setlength\rowgap{8pt}%
  \newcommand\pad[1]{\raisebox{0pt}[\dimexpr\height+\rowgap\relax][\dimexpr\depth+\rowgap\relax]{\strut#1}}%
\begin{table}
\def\arraystretch{1}
\setlength{\tabcolsep}{10pt}
\caption{\label{table:app:vacuum-state:E_n4_u} Lookout table for the coefficients beside $\ket{u}$}
\begin{tabular}{c|cccl}
    $\ket{u}$  & $E^u_{n \le4}$                                                     & $n$ & partition of $n$ & \pad{Type of state} \\ \hline
    $\ket{1}$  & Equation~\eqref{eq:app:vacuum-state:eigenstate_condition_n_4_u_1}  & $1$ & $(1)$            & \pad{$\ket{S_{\mu,\alpha_1}^{z_{\alpha_1}} = S_{\alpha_1}-1}$} \\
    $\ket{2}$  & Equation~\eqref{eq:app:vacuum-state:eigenstate_condition_n_4_u_2}  & $2$ & $(2+0)$          & \pad{$\ket{S_{\mu,\alpha_1}^{z_{\alpha_1}} = S_{\alpha_1}-2}$} \\
    $\ket{3}$  & Equation~\eqref{eq:app:vacuum-state:eigenstate_condition_n_4_u_3}  & $2$ & $(1+1)$          & \pad{$\Ket{\begin{aligned}&S_{\mu,\alpha_1}^{z_{\alpha_1}} = S_{\alpha_1}-1;\\ &S_{\mu+\nu_2,\alpha_2}^{z_{\alpha_2}} = S_{\alpha_2}-1\end{aligned}}$} \\
    $\ket{4}$  & Equation~\eqref{eq:app:vacuum-state:eigenstate_condition_n_4_u_4}  & $3$ & $(3+0+0)$        & \pad{$\ket{S_{\mu,\alpha_1}^{z_{\alpha_1}} = S_{\alpha_1}-3}$} \\
    $\ket{5}$  & Equation~\eqref{eq:app:vacuum-state:eigenstate_condition_n_4_u_5}  & $3$ & $(2+1+0)$        & \pad{$\Ket{\begin{aligned}&S_{\mu,\alpha_1}^{z_{\alpha_1}} = S_{\alpha_1}-2; \\ &S_{\mu+\nu_2,\alpha_2}^{z_{\alpha_2}} = S_{\alpha_2}-1\end{aligned}}$} \\
    $\ket{6}$  & Equation~\eqref{eq:app:vacuum-state:eigenstate_condition_n_4_u_6}  & $3$ & $(1+1+1)$        & \pad{$\Ket{\begin{aligned}&S_{\mu,\alpha_1}^{z_{\alpha_1}} = S_{\alpha_1}-1; \\ &S_{\mu+\nu_2,\alpha_2}^{z_{\alpha_2}} = S_{\alpha_2}-1;\\ &S_{\mu+\nu_3,\alpha_3}^{z_{\alpha_3}} = S_{\alpha_3}-1\end{aligned}}$} \\
    $\ket{7}$  & Equation~\eqref{eq:app:vacuum-state:eigenstate_condition_n_4_u_7}  & $4$ & $(4+0+0+0)$      & \pad{$\ket{S_{\mu,\alpha_1}^{z_{\alpha_1}} = S_{\alpha_1}-4}$} \\
    $\ket{8}$  & Equation~\eqref{eq:app:vacuum-state:eigenstate_condition_n_4_u_8}  & $4$ & $(3+1+0+0)$      & \pad{$\Ket{\begin{aligned}&S_{\mu,\alpha_1}^{z_{\alpha_1}} = S_{\alpha_1}-3;\\ &S_{\mu+\nu_2,\alpha_2}^{z_{\alpha_2}} = S_{\alpha_2}-1\end{aligned}}$} \\
    $\ket{9}$  & Equation~\eqref{eq:app:vacuum-state:eigenstate_condition_n_4_u_9}  & $4$ & $(2+2+0+0)$      & \pad{$\Ket{\begin{aligned}&S_{\mu,\alpha_1}^{z_{\alpha_1}} = S_{\alpha_1}-2;\\ &S_{\mu+\nu_2,\alpha_2}^{z_{\alpha_2}} = S_{\alpha_2}-2\end{aligned}}$} \\
    $\ket{10}$ & Equation~\eqref{eq:app:vacuum-state:eigenstate_condition_n_4_u_10} & $4$ & $(2+1+1+0)$      & \pad{$\Ket{\begin{aligned}&S_{\mu,\alpha_1}^{z_{\alpha_1}} = S_{\alpha_1}-2;\\ &S_{\mu+\nu_2,\alpha_2}^{z_{\alpha_2}} = S_{\alpha_2}-1;\\ &S_{\mu+\nu_3,\alpha_2}^{z_{\alpha_3}} = S_{\alpha_3}-1\end{aligned}}$} \\
    $\ket{11}$ & Equation~\eqref{eq:app:vacuum-state:eigenstate_condition_n_4_u_11} & $4$ & $(1+1+1+1)$      & \pad{$\Ket{\begin{aligned}&S_{\mu,\alpha_1}^{z_{\alpha_1}} = S_{\alpha_1}-1; \\ &S_{\mu+\nu_2,\alpha_2}^{z_{\alpha_2}} = S_{\alpha_2}-1;\\ &S_{\mu+\nu_3,\alpha_3}^{z_{\alpha_3}} = S_{\alpha_3}-1;\\ &S_{\mu+\nu_4,\alpha_4}^{z_{\alpha_4}} = S_{\alpha_4}-1\end{aligned}}$} \\
\end{tabular}
\end{table}
}

\clearpage
\begin{eqnarray}
    E_{n\le 4}^{u=1} &=& 
    \dfrac{\sqrt{2S_{\alpha_1}}}{2}
    \cdot\nonumber\\&&\cdot
    \Biggl(
        \sum_{i_1}
        \tilde{J}_{\alpha_1}^{i_1}
        p^{i_1}_{\alpha_1}
        -
        \sum_{i_1, i_2}
        \tilde{J}_{0;\alpha_1,\alpha_1}^{i_1,i_2}
        z^{i_1}_{\alpha_1}
        p^{i_2}_{\alpha_1}
        +\nonumber\\&&\phantom{\cdot\Biggl(}+
        \sum_{i_1,i_2,i_3}
        \tilde{J}_{0,0;\alpha_1,\alpha_1,\alpha_1}^{i_1,i_2,i_3}
        \Biggl(
            z_{\alpha_1}^{i_1} z_{\alpha_1}^{i_2}
            +
            \frac{1}{2} S_{\alpha_1} p_{\alpha_1}^{i_1} \overline{p_{\alpha_1}^{i_2}}
            +
            \left(S_{\alpha_1}-\frac{1}{2}\right) \overline{p_{\alpha_1}^{i_1}} p_{\alpha_1}^{i_2}
        \Biggr)
        p_{\alpha_1}^{i_3}
        +\nonumber\\&&\phantom{\cdot\Biggl(}+
        \sum_{\nu_2,\alpha_2}
        [\bm{r}_{\mu,\alpha_1} \ne \bm{r}_{\mu+\nu_2,\alpha_2}]
        \sum_{i_1,i_2,i_3}
        \dfrac{3}{2} 
        S_{\alpha_2}
        \tilde{J}_{\nu_2,\nu_2;\alpha_1,\alpha_2,\alpha_2}^{i_1,i_2,i_3}
        p_{\alpha_1}^{i_1}
        \overline{p_{\alpha_2}^{i_2}}
        p_{\alpha_2}^{i_3}
        -\nonumber\\&&\phantom{\cdot\Biggl(}-
        \sum_{\substack{i_1,i_2,\\i_3,i_4}}
        \tilde{J}_{0,0,0;\alpha_1\alpha_1,\alpha_1\alpha_1}^{i_1,i_2,i_3,i_4}
        \Biggl(
            z_{\alpha_1}^{i_1}
            z_{\alpha_1}^{i_2}
            z_{\alpha_1}^{i_3}
            +
            \frac{1}{2}S_{\alpha_1}
            \Bigl(
                p_{\alpha_1}^{i_1}
                p_{\alpha_1}^{i_2}
                z_{\alpha_1}^{i_3}
                +
                z_{\alpha_1}^{i_1}
                p_{\alpha_1}^{i_2}
                p_{\alpha_1}^{i_3}
            \Bigr)
            +\nonumber\\&&\phantom{\cdot\Biggl(-\sum_{\substack{i_1,i_2,\\i_3,i_4}}\Biggl[\tilde{J}_{0,0,0;\alpha_1\alpha_1,\alpha_1\alpha_1}^{i_1,i_2,i_3,i_4}\Biggl(}+
            \left(S_{\alpha_1}-\frac{1}{2}\right)
            \Bigl(
                2\overline{p_{\alpha_1}^{i_1}}
                z_{\alpha_1}^{i_2}
                p_{\alpha_1}^{i_3}
                +
                z_{\alpha_1}^{i_1}
                \overline{p_{\alpha_1}^{i_2}}
                p_{\alpha_1}^{i_3}
                +
                \overline{p_{\alpha_1}^{i_1}}
                p_{\alpha_1}^{i_2}
                z_{\alpha_1}^{i_3}
            \Bigr)
        \Biggr)
        p_{\alpha_1}^{i_4}
        +\nonumber\\&&\phantom{\cdot\Biggl(}+
        \sum_{\nu_2,\alpha_2}
        [\bm{r}_{\mu,\alpha_1} \ne \bm{r}_{\mu+\nu_2,\alpha_2}]
        \sum_{\substack{i_1,i_2,\\i_3,i_4}}
        2S_{\alpha_2}
        \tilde{J}_{\nu_2,\nu_2,\nu_2;\alpha_1\alpha_2,\alpha_2\alpha_2}^{i_1,i_2,i_3,i_4}
        p_{\alpha_1}^{i_1}
        \overline{p_{\alpha_2}^{i_2}}
        z_{\alpha_2}^{i_3}
        p_{\alpha_2}^{i_4}
        +\nonumber\\&&\phantom{\cdot\Biggl(}+
        \sum_{\nu_2,\alpha_2}
        [\bm{r}_{\mu,\alpha_1} \ne \bm{r}_{\mu+\nu_2,\alpha_2}]
        \sum_{\substack{i_1,i_2,\\i_3,i_4}}
        3S_{\alpha_2}
        \tilde{J}_{0,\nu_2,\nu_2;\alpha_1\alpha_1,\alpha_2\alpha_2}^{i_1,i_2,i_3,i_4}
        z_{\alpha_1}^{i_1}
        p_{\alpha_1}^{i_2}
        \overline{p_{\alpha_2}^{i_3}}
        p_{\alpha_2}^{i_4}
    \Biggr).
    \label{eq:app:vacuum-state:eigenstate_condition_n_4_u_1}
    \\
    E_{n\le 4}^{u=2}
    &=& 
    \dfrac{\sqrt{S_{\alpha_1}(2S_{\alpha_1}-1)}}{2}
    \cdot\nonumber\\&&\cdot
    \Biggl(
        \sum_{i_1,i_2}
        \tilde{J}_{0;\alpha_1,\alpha_1}^{i_1,i_2}
        p_{\alpha_1}^{i_1}
        p_{\alpha_1}^{i_2}
        -
        \sum_{i_1,i_2,i_3}
        \tilde{J}_{0,0;\alpha_1,\alpha_1,\alpha_1}^{i_1,i_2,i_3}
        \Bigl(
            2 z_{\alpha_1}^{i_1} p_{\alpha_1}^{i_2}
            +
            p_{\alpha_1}^{i_1} z_{\alpha_1}^{i_2} 
        \Bigr)
        p_{\alpha_1}^{i_3}
        +\nonumber\\&&\phantom{\cdot\Biggl(}+
        \sum_{\substack{i_1,i_2,\\i_3,i_4}}
        \tilde{J}_{0,0,0;\alpha_1,\alpha_1,\alpha_1,\alpha_1}^{i_1,i_2,i_3,i_4}
        \Biggl(
            4z_{\alpha_1}^{i_1}
            z_{\alpha_1}^{i_2}
            p_{\alpha_1}^{i_3}
            +
            2z_{\alpha_1}^{i_1}
            p_{\alpha_1}^{i_2}
            z_{\alpha_1}^{i_3}
            +
            p_{\alpha_1}^{i_1}
            z_{\alpha_1}^{i_2}
            z_{\alpha_1}^{i_3}
            +\nonumber\\&&\phantom{\cdot\Biggl(+\sum_{\substack{i_1,i_2,\\i_3,i_4}}\tilde{J}_{0,0,0;\alpha_1,\alpha_1,\alpha_1,\alpha_1}^{i_1,i_2,i_3,i_4}\Biggl(}+
            \frac{1}{2} S_{\alpha_1}
            p_{\alpha_1}^{i_1}
            p_{\alpha_1}^{i_2}
            \overline{p_{\alpha_1}^{i_3}}
            +
            \left(S_{\alpha_1}-\frac{1}{2}\right)
            p_{\alpha_1}^{i_1}
            \overline{p_{\alpha_1}^{i_2}}
            p_{\alpha_1}^{i_3}
            +\nonumber\\&&\phantom{\cdot\Biggl(+\sum_{\substack{i_1,i_2,\\i_3,i_4}}\tilde{J}_{0,0,0;\alpha_1,\alpha_1,\alpha_1,\alpha_1}^{i_1,i_2,i_3,i_4}\Biggl(}+
            \frac{3}{2}\left(S_{\alpha_1}-1\right)
            \overline{p_{\alpha_1}^{i_1}}
            p_{\alpha_1}^{i_2}
            p_{\alpha_1}^{i_3}
        \Biggr)
        p_{\alpha_1}^{i_4}
        +\nonumber\\&&\phantom{\cdot\Biggl(}+
        \sum_{\nu_2,\alpha_2}
        [\bm{r}_{\mu,\alpha_1} \ne \bm{r}_{\mu+\nu_2,\alpha_2}]
        \sum_{\substack{i_1,i_2,\\i_3,i_4}}
        3S_{\alpha_2}
        \tilde{J}_{0,\nu_2,\nu_2;\alpha_1,\alpha_1,\alpha_2,\alpha_2}^{i_1,i_2,i_3,i_4}
        p_{\alpha_1}^{i_1}
        p_{\alpha_1}^{i_2}
        \overline{p_{\alpha_2}^{i_3}}
        p_{\alpha_2}^{i_4}
    \Biggr).
    \label{eq:app:vacuum-state:eigenstate_condition_n_4_u_2}
\end{eqnarray}

\newpage
\begin{eqnarray}
    E_{n\le 4}^{u=3}
    &=&
    \dfrac{\sqrt{S_{\alpha_1}S_{\alpha_2}}}{2}
    [\bm{r}_{\mu,\alpha_1} \ne \bm{r}_{\mu+\nu_2,\alpha_2}]
    \cdot\nonumber\\&&\cdot
    \Biggl(
        \sum_{i_1,i_2}
        \tilde{J}_{\nu_2;\alpha_1,\alpha_2}^{i_1,i_2}
        p_{\alpha_1}^{i_1}
        p_{\alpha_2}^{i_2}
        -
        \sum_{i_1,i_2,i_3}
        3\tilde{J}_{\nu_2,\nu_2;\alpha_1,\alpha_2,\alpha_2}^{i_1,i_2,i_3}
        p_{\alpha_1}^{i_1}
        z_{\alpha_2}^{i_2}
        p_{\alpha_2}^{i_3}
        +\nonumber\\&&\phantom{\cdot\Biggl(}+
        \sum_{\substack{i_1,i_2,\\i_3,i_4}}
        4\tilde{J}_{0,0,\nu_2;\alpha_1,\alpha_1,\alpha_1,\alpha_2}^{i_1,i_2,i_3,i_4}
        \Biggl(
            z_{\alpha_1}^{i_1}
            z_{\alpha_1}^{i_2}
            +
            \left(S_{\alpha_1}-\frac{1}{2}\right)
            \overline{p_{\alpha_1}^{i_1}}
            p_{\alpha_1}^{i_2}
            +
            \dfrac{S_{\alpha_1}}{2}
            p_{\alpha_1}^{i_1}
            \overline{p_{\alpha_1}^{i_2}}
        \Biggr)
        p_{\alpha_1}^{i_3}
        p_{\alpha_2}^{i_4}
        +\nonumber\\&&\phantom{\cdot\Biggl(}+
        \sum_{\substack{i_1,i_2,\\i_3,i_4}}
        3\tilde{J}_{0,\nu_2,\nu_2;\alpha_1,\alpha_1,\alpha_2,\alpha_2}^{i_1,i_2,i_3,i_4}
        z_{\alpha_1}^{i_1}
        p_{\alpha_1}^{i_2}
        z_{\alpha_2}^{i_3}
        p_{\alpha_2}^{i_4}
        +\nonumber\\&&\phantom{\cdot\Biggl(}+
        \sum_{\nu_3,\alpha_3}
        [\bm{r}_{\mu,\alpha_1} \ne \bm{r}_{\mu+\nu_3,\alpha_3}]
        [\bm{r}_{\mu+\nu_2,\alpha_2} \ne \bm{r}_{\mu+\nu_3,\alpha_3}]
        \cdot\nonumber\\&&\phantom{\cdot\Biggl(+\sum_{\nu_3,\alpha_3}}\cdot
        \sum_{\substack{i_1,i_2,\\i_3,i_4}}
        3S_{\alpha_3}
        \tilde{J}_{\nu_2,\nu_3,\nu_3;\alpha_1,\alpha_2,\alpha_3,\alpha_3}^{i_1,i_2,i_3,i_4}
        p_{\alpha_1}^{i_1}
        p_{\alpha_2}^{i_2}
        \overline{p_{\alpha_3}^{i_3}}
        p_{\alpha_3}^{i_4}
    \Biggr).
    \label{eq:app:vacuum-state:eigenstate_condition_n_4_u_3}
\\
    E_{n\le 4}^{u=4}
    &=&
    \dfrac{\sqrt{3S_{\alpha_1}\left(S_{\alpha_1}-\frac{1}{2}\right)(S_{\alpha_1}-1)}}{2}
    \cdot\nonumber\\&&\cdot
    \Biggl(
        \sum_{i_1,i_2,i_3}
        \tilde{J}_{0,0;\alpha_1,\alpha_1,\alpha_1}^{i_1,i_2,i_3}
        p_{\alpha_1}^{i_1}
        p_{\alpha_1}^{i_2}
        p_{\alpha_1}^{i_3}
        -\nonumber\\&&\phantom{\cdot\Biggl(}-
        \sum_{\substack{i_1,i_2,\\i_3,i_4}}
        \tilde{J}_{0,0,0;\alpha_1,\alpha_1,\alpha_1,\alpha_1}^{i_1,i_2,i_3,i_4}
        \Bigl(
            3z_{\alpha_1}^{i_1}
            p_{\alpha_1}^{i_2}
            p_{\alpha_1}^{i_3}
            +
            2p_{\alpha_1}^{i_1}
            z_{\alpha_1}^{i_2}
            p_{\alpha_1}^{i_3}
            +
            p_{\alpha_1}^{i_1}
            p_{\alpha_1}^{i_2}
            z_{\alpha_1}^{i_3}
        \Bigr)
        p_{\alpha_1}^{i_4}
    \Biggr).
    \label{eq:app:vacuum-state:eigenstate_condition_n_4_u_4}
\\
    E_{n\le 4}^{u=5}
    &=&
    \dfrac{\sqrt{S_{\alpha_1}\left(S_{\alpha_1}-\frac{1}{2}\right)S_{\alpha_2}}}{2}
    [\bm{r}_{\mu,\alpha_1} \ne \bm{r}_{\mu+\nu_2,\alpha_2}]
    \cdot\nonumber\\&&\cdot
    \Biggl(
        \sum_{i_1,i_2,i_3}
        3\tilde{J}_{0,\nu_2;\alpha_1,\alpha_1,\alpha_2}^{i_1,i_2,i_3}
        p_{\alpha_1}^{i_1}
        p_{\alpha_1}^{i_2}
        p_{\alpha_2}^{i_3}
        -\nonumber\\&&\phantom{\cdot\Biggl(}-
        \sum_{\substack{i_1,i_2,\\i_3,i_4}}
        4\tilde{J}_{0,0,\nu_2;\alpha_1,\alpha_1,\alpha_1,\alpha_2}^{i_1,i_2,i_3,i_4}
        \Bigl(
            2z_{\alpha_1}^{i_1}
            p_{\alpha_1}^{i_2}
            +
            p_{\alpha_1}^{i_1}
            z_{\alpha_1}^{i_2}
        \Bigr)
        p_{\alpha_1}^{i_3}
        p_{\alpha_2}^{i_4}
        -\nonumber\\&&\phantom{\cdot\Biggl(}-
        \sum_{\substack{i_1,i_2,\\i_3,i_4}}
        6\tilde{J}_{0,\nu_2,\nu_2;\alpha_1,\alpha_1,\alpha_2,\alpha_2}^{i_1,i_2,i_3,i_4}
        p_{\alpha_1}^{i_1}
        p_{\alpha_1}^{i_2}
        z_{\alpha_2}^{i_3}
        p_{\alpha_2}^{i_4}
    \Biggr).
    \label{eq:app:vacuum-state:eigenstate_condition_n_4_u_5}
\\
    E_{n\le 4}^{u=6}
    &=&
    \dfrac{\sqrt{2S_{\alpha_1}S_{\alpha_2}S_{\alpha_3}}}{4}
    [\bm{r}_{\mu,\alpha_1}, \bm{r}_{\mu+\nu_2,\alpha_2}, \bm{r}_{\mu+\nu_3,\alpha_3}\text{ are all different}]
    \cdot\nonumber\\&&\cdot
    \Biggl(
        \sum_{i_1,i_2,i_3}
        \tilde{J}_{\nu_2,\nu_3;\alpha_1,\alpha_2,\alpha_3}^{i_1,i_2,i_3}
        p_{\alpha_1}^{i_1}
        p_{\alpha_2}^{i_2}
        p_{\alpha_3}^{i_3}
        -
        \sum_{\substack{i_1,i_2,\\i_3,i_4}}
        6\tilde{J}_{\nu_2,\nu_3,\nu_3;\alpha_1,\alpha_2,\alpha_3,\alpha_3}^{i_1,i_2,i_3,i_4}
        p_{\alpha_1}^{i_1}
        p_{\alpha_2}^{i_2}
        z_{\alpha_3}^{i_3}
        p_{\alpha_3}^{i_4}
    \Biggr).
    \label{eq:app:vacuum-state:eigenstate_condition_n_4_u_6}
\end{eqnarray}

\newpage
\begin{eqnarray}
    E_{n\le 4}^{u=7}
    &=&
    \dfrac{\sqrt{6S_{\alpha_1}\left(S_{\alpha_1}-\frac{1}{2}\right)\left(S_{\alpha_1}-1\right)\left(S_{\alpha_1}-\frac{3}{2}\right)}}{2}
    \cdot\nonumber\\&&\cdot
    \sum_{\substack{i_1,i_2,\\i_3,i_4}}
    \tilde{J}_{0,0,0;\alpha_1,\alpha_1,\alpha_1,\alpha_1}^{i_1,i_2,i_3,i_4}
    p_{\alpha_1}^{i_1}
    p_{\alpha_1}^{i_2}
    p_{\alpha_1}^{i_3}
    p_{\alpha_1}^{i_4}.
    \label{eq:app:vacuum-state:eigenstate_condition_n_4_u_7}
\\
    E_{n\le 4}^{u=8}
    &=&
    \dfrac{\sqrt{6S_{\alpha_1}\left(S_{\alpha_1}-\frac{1}{2}\right)\left(S_{\alpha_1}-1\right)S_{\alpha_2}}}{4}
    [\bm{r}_{\mu,\alpha_1} \ne \bm{r}_{\mu+\nu_2,\alpha_2}]
    \cdot\nonumber\\&&\cdot
    \sum_{\substack{i_1,i_2,\\i_3,i_4}}
    7\tilde{J}_{0,0,\nu_2;\alpha_1,\alpha_1,\alpha_1,\alpha_2}^{i_1,i_2,i_3,i_4}
    p_{\alpha_1}^{i_1}
    p_{\alpha_1}^{i_2}
    p_{\alpha_1}^{i_3}
    p_{\alpha_2}^{i_4}.
    \label{eq:app:vacuum-state:eigenstate_condition_n_4_u_8}
\\
    E_{n\le 4}^{u=9}
    &=&
    \dfrac{\sqrt{S_{\alpha_1}\left(S_{\alpha_1}-\frac{1}{2}\right)S_{\alpha_2}\left(S_{\alpha_2}-\frac{1}{2}\right)}}{2}
    [\bm{r}_{\mu,\alpha_1} \ne \bm{r}_{\mu+\nu_2,\alpha_2}]
    \cdot\nonumber\\&&\cdot
    \sum_{\substack{i_1,i_2,\\i_3,i_4}}
    3\tilde{J}_{0,\nu_2,\nu_2;\alpha_1,\alpha_1,\alpha_2,\alpha_2}^{i_1,i_2,i_3,i_4}
    p_{\alpha_1}^{i_1}
    p_{\alpha_1}^{i_2}
    p_{\alpha_2}^{i_3}
    p_{\alpha_2}^{i_4}.
    \label{eq:app:vacuum-state:eigenstate_condition_n_4_u_9}
\\
    E_{n\le 4}^{u=10}
    &=&
    \dfrac{\sqrt{S_{\alpha_1}\left(S_{\alpha_1}-\frac{1}{2}\right)S_{\alpha_2}S_{\alpha_3}}}{2}
    [\bm{r}_{\mu,\alpha_1}, \bm{r}_{\mu+\nu_2,\alpha_2}, \bm{r}_{\mu+\nu_3,\alpha_3}\text{ are all different}]
    \cdot\nonumber\\&&\cdot
    \sum_{\substack{i_1,i_2,\\i_3,i_4}}
    3\tilde{J}_{0,\nu_2,\nu_3;\alpha_1,\alpha_1,\alpha_2,\alpha_3}^{i_1,i_2,i_3,i_4}
    p_{\alpha_1}^{i_1}
    p_{\alpha_1}^{i_2}
    p_{\alpha_2}^{i_3}
    p_{\alpha_3}^{i_4}.
    \label{eq:app:vacuum-state:eigenstate_condition_n_4_u_10}
\\
    E_{n\le 4}^{u=11}
    &=&
    \dfrac{\sqrt{S_{\alpha_1}S_{\alpha_2}S_{\alpha_3}S_{\alpha_4}}}{2}
    [\bm{r}_{\mu,\alpha_1}, \bm{r}_{\mu+\nu_2,\alpha_2}, \bm{r}_{\mu+\nu_3,\alpha_3}, \bm{r}_{\mu+\nu_4,\alpha_4}\text{ are all different}]
    \cdot\nonumber\\&&\cdot
    \sum_{\substack{i_1,i_2,\\i_3,i_4}}
    3\tilde{J}_{\nu_2,\nu_3,\nu_4;\alpha_1,\alpha_2,\alpha_3,\alpha_4}^{i_1,i_2,i_3,i_4}
    p_{\alpha_1}^{i_1}
    p_{\alpha_2}^{i_2}
    p_{\alpha_3}^{i_3}
    p_{\alpha_4}^{i_4}.
    \label{eq:app:vacuum-state:eigenstate_condition_n_4_u_11}
\end{eqnarray}

Finally, we note that in the case $n \le 2$, which can often be found in the literature, only the states $\ket{1}$, $\ket{2}$, $\ket{3}$ have non-zero coefficients beside them. Therefore, the eigenstate condition simplifies to $E_{n\le 2}^{u} = 0$, where $u = 1, 2, 3$ and $E_{n\le 2}^u$ is a simplified form of $E_{n\le 4}^u$, with all higher order parameters being set to zero. These conditions are tabulated below for reference

\begin{subequations}
\begin{eqnarray}
    E_{n\le 2}^{u=1} = 0
    &\Rightarrow&
    \sum_{i_1}
    \tilde{J}_{\alpha_1}^{i_1}
    p^{i_1}_{\alpha_1}
    -
    \sum_{i_1, i_2}
    \tilde{J}_{0;\alpha_1,\alpha_1}^{i_1,i_2}
    z^{i_1}_{\alpha_1}
    p^{i_2}_{\alpha_1}
    =0
    \\
    E_{n\le 2}^{u=2} = 0
    &\Rightarrow&
    \sqrt{2S_{\alpha_1}-1}
    \sum_{i_1,i_2}
    \tilde{J}_{0;\alpha_1,\alpha_1}^{i_1,i_2}
    p^{i_1}_{\alpha_1}
    p^{i_2}_{\alpha_1}
    =0
    \\
    E_{n\le 2}^{u=3} = 0
    &\Rightarrow&
    [\bm{r}_{\mu,\alpha_1} \ne \bm{r}_{\mu+\nu_2,\alpha_2}]
    \sum_{i_1,i_2}
    \tilde{J}_{\nu_2;\alpha_1,\alpha_2}^{i_1,i_2}
    p^{i_1}_{\alpha_1}
    p^{i_2}_{\alpha_2}
    =0
\end{eqnarray}
\label{eq:app:vacuum-state:eigenstate_condition_n_2_u_123}
\end{subequations}

\newpage
\section{\label{app:renormalization}Renormalization of parameters}

Here we summarize the renormalization procedure for the interaction parameters (i.~e. transition from Equation~\EqAllMightyHamiltonian to Equation~\EqRenormalizedAllMightyHamiltonian~and~\EqRenormalizedParameters of the main text).

\subsection{Renormalization procedure}

By inserting the representation of the spin operator from Equation~\EqSpinOperatorDeviation into the Hamiltonian in Equation~\EqAllMightyHamiltonian of the main text, we get
\begin{equation}
    \mathcal{\hat{H}}
    =
    \sum_{m=1}^{\infty}
    C_m
    \sum_{\substack{\mu, \nu_2,\dots,\nu_m,\\ \alpha_1,\dots,\alpha_m,\\ i_1,\dots,i_m}}
    \Biggl(
        J^{i_1,\dots,i_n}_{\nu_2,\dots,\nu_m;\alpha_1,\dots,\alpha_m}
        \prod_{l=1}^m \left(S_{\alpha_l}z^{i_l}_{\alpha_l} + \hat{\Delta}^{i_l}_{\mu+\nu_l,\alpha_l}\right)
    \Biggr).
    \label{eq:app:renormalization:deviation_repr_inserted_in_hamiltonian}
\end{equation}

Let us select an arbitrary value of $m$, that we denote as $n$. We can then separate the sum over $m$ in Equation~\eqref{eq:app:renormalization:deviation_repr_inserted_in_hamiltonian} into two groups as
\begin{equation}
    \sum_{m=1}^{\infty} \ldots
    =
    \sum_{m=1}^{m=n-1}\ldots
    +
    \sum_{m=n}^{\infty} \ldots.
    \label{eq:app:renormalization:split_of_the_sum}
\end{equation}

The terms 
\begin{equation}
    \sum_{\substack{\mu, \nu_2\dots,\nu_n,\\ \alpha_1,\dots,\alpha_n,\\ i_1,\dots,i_n}}
    \Biggl(
        \tilde{J}^{i_1,\dots,i_n}_{\nu_2,\dots,\nu_n;\alpha_1,\dots,\alpha_n}
        \prod_{m=1}^{n} \hat{\Delta}^{i_m}_{\mu+\nu_m,\alpha_m}
    \Biggr)
    \label{eq:app:renormalization:renormalized-term-of-interest}
\end{equation}
of Equation~\EqRenormalizedAllMightyHamiltonian of the main text can originate only from the second sum over $n$ in Equation~\eqref{eq:app:renormalization:split_of_the_sum} as every term of the first one contains at most $n-1$ deviation operators. In other words, renormalized parameters $\tilde{J}^{i_1,\dots,i_n}_{\nu_2,\dots,\nu_n;\alpha_1,\dots,\alpha_n}$ contain contributions from the original parameters $J^{i_1,\dots,i_m}_{\nu_2,\dots,\nu_m;\alpha_1,\dots,\alpha_m}$ only if $m \ge n$. 

At this point we ask ourself the question: how many terms of the initial Hamiltonian (Equation~\EqAllMightyHamiltonian of the main text) with $J^{i_1,\dots,i_m}_{\nu_2,\dots,\nu_m;\alpha_1,\dots,\alpha_m}$ in them contribute to the term of the renormalized Hamiltonian (Equation~\EqRenormalizedAllMightyHamiltonian of the main text) with $\tilde{J}^{i_1,\dots,i_n}_{\nu_2,\dots,\nu_n;\alpha_1,\dots,\alpha_n}$ in it? The answer to this question is given by the binomial coefficient
\begin{equation}
    \begin{pmatrix}
        m\\n
    \end{pmatrix}
    =
    \dfrac{m!}{n!(m-n)!}.
    \label{eq:app:renormalization:binomial-coefficient}
\end{equation}
since the renormalized parameter is multiplied by $n$ deviation operators and each binomial $\left(S_{\alpha_l}z_{\alpha_l}^{i_l} + \hat{\Delta}_{\mu+\nu_l,\alpha_l}^{i_l}\right)$ can contribute at most one deviation operator when the product $\prod_{l=1}^m$ is expanded in Equation~\eqref{eq:app:renormalization:deviation_repr_inserted_in_hamiltonian}.
We can then construct the renormalized parameters for any given value of $n$ and $m$. In the special case when positions of sites associated with both initial and renormalized parameters are different or for $n,m \le 2$ and arbitrary positions, the renormalized parameters are given by a simple formula
\begin{equation}
    \tilde{J}^{i_1,\dots,i_n}_{\nu_2,\dots,\nu_n;\alpha_1,\dots,\alpha_n}
    =
    \sum_{m=n}^{\infty}
    C_m
    \dfrac{m!}{n!(m-n)!}
    \sum_{\substack{\nu_{n+1},\dots,\nu_m,\\ \alpha_{n+1},\dots,\alpha_m,\\ i_{n+1},\dots,i_m}}
    \Biggl(
        J^{i_1,\dots,i_m}_{\nu_2,\dots,\nu_m;\alpha_1,\dots,\alpha_m}
        \prod_{l=n+1}^m S_{\alpha_l}z^{i_l}_{\alpha_l}
    \Biggr)
    \label{eq:app:renormalization:renormalized-parameters-all-diff}
\end{equation}
in which we assume symmetric gauge for the initial parameters as described in~\ref{app:gauge}.

Unfortunately, when some of the sites are located at the same position in real space, the spin operators do not commute and no general expression can be written for the renormalized parameters. In this case we introduce the sum over the special index $p_{n,m}$. This sum always contains the $m!/(n!(m-n)!)$ terms and each value of the index $p_{n,m}$ is associated with the codependent permutation of the Cartesian and position indices, which we denote as $p_{n,m}(i_1, \ldots, i_m)$ and $p_{n,m}(\nu_2,\ldots,\nu_m;\alpha_1,\ldots,\alpha_m)$ respectively. These permutations are constructed for each pair of $n,m$ by expanding the product of binomials in Equation~\eqref{eq:app:renormalization:deviation_repr_inserted_in_hamiltonian} for a fixed $m$, taking the terms with $n$ spin deviation operators and renaming the indices in such a way that all those terms are multiplied by $\prod_{l=1}^n \Delta_{\alpha_l}z^{i_l}_{\alpha_l}\prod_{l=n+1}^m S_{\alpha_l}z^{i_l}_{\alpha_l}$. We use this new notation to write the Equation~\EqRenormalizedParameters of the main text.

Below we list the renormalized parameters for $n,m \le 4$. We note that the Equation~\eqref{eq:app:renormalization:renormalized-params-n-1-m-le-4}~--~\eqref{eq:app:renormalization:renormalized-params-n-4-m-le-4} can be partially simplified toward the form in Equation~\eqref{eq:app:renormalization:renormalized-parameters-all-diff} by using the hermiticity of the initial Hamiltonian (described in~\ref{app:hermiticity}) and the choice of symmetric gauge of initial parameters (described in~\ref{app:gauge}). However, that "simplified" form is unnecessary convoluted and do not provide any value to our discussion. Thus, we use Equation~\eqref{eq:app:renormalization:renormalized-params-n-1-m-le-4}~--~\eqref{eq:app:renormalization:renormalized-params-n-4-m-le-4} as is. They are valid for any combination of site's positions and are the unique equations in this article that are free from the assumption of hermiticity and from the choice of gauge

\begin{eqnarray}
    \tilde{J}^{i_1}_{\alpha_1}
    &=&
    C_1
    J^{i_1}_{\alpha_1}
    +
    C_2
    \sum_{\substack{\nu_2,\\\alpha_2,\\i_2}}
    \Biggl(
        J^{i_1,i_2}_{\nu_2;\alpha_1,\alpha_2}
        +
        J^{i_2,i_1}_{-\nu_2;\alpha_2,\alpha_1}
    \Biggr)
    S_{\alpha_2}z^{i_2}_{\alpha_2}
    +\nonumber\\&&+
    C_3
    \sum_{\substack{\nu_2,\nu_3,\\\alpha_2,\alpha_3,\\i_2,i_3}}
    \Biggl(
        J^{i_1,i_2,i_3}_{\nu_2,\nu_3,\alpha_1,\alpha_2,\alpha_3}
        +
        J^{i_2,i_1,i_3}_{-\nu_2,\nu_3-\nu_2,\alpha_2,\alpha_1,\alpha_3}
        +\nonumber\\&&\phantom{+C_3\sum_{\substack{\nu_2,\nu_3,\\\alpha_2,\alpha_3,\\i_2,i_3}}\Biggl(J^{i_1,i_2,i_3}_{\nu_2,\nu_3,\alpha_1,\alpha_2,\alpha_3}}+
        J^{i_3,i_2,i_1}_{\nu_2-\nu_3,-\nu_3,\alpha_3,\alpha_2,\alpha_1}
    \Biggr)
    S_{\alpha_2}z^{i_2}_{\alpha_2}
    S_{\alpha_3}z^{i_3}_{\alpha_3}
    +\nonumber\\&&+
    C_4
    \sum_{\substack{\nu_2,\nu_3,\nu_4\\\alpha_2,\alpha_3,\alpha_4\\i_2,i_3,i_4}}
    \Biggl(
        J^{i_1,i_2,i_3,i_4}_{\nu_2,\nu_3,\nu_4;\alpha_1,\alpha_2,\alpha_3,\alpha_4}
        +\nonumber\\&&\phantom{+C_4\sum_{\substack{\nu_2,\nu_3,\nu_4\\\alpha_2,\alpha_3,\alpha_4\\i_2,i_3,i_4}}\Bigl(}+
        J^{i_2,i_1,i_3,i_4}_{-\nu_2,\nu_3-\nu_2,\nu_4-\nu_2;\alpha_2,\alpha_1,\alpha_3,\alpha_4}
        +\nonumber\\&&\phantom{+C_4\sum_{\substack{\nu_2,\nu_3,\nu_4\\\alpha_2,\alpha_3,\alpha_4\\i_2,i_3,i_4}}\Bigl(}+
        J^{i_3,i_2,i_1,i_4}_{\nu_2-\nu_3,-\nu_3,\nu_4-\nu_3;\alpha_3,\alpha_2,\alpha_1,\alpha_4}
        +\nonumber\\&&\phantom{+C_4\sum_{\substack{\nu_2,\nu_3,\nu_4\\\alpha_2,\alpha_3,\alpha_4\\i_2,i_3,i_4}}\Bigl(}+
        J^{i_4,i_2,i_3,i_1}_{\nu_2-\nu_4,\nu_3-\nu_4,-\nu_4;\alpha_4,\alpha_2,\alpha_3,\alpha_1}
    \Biggr)
    S_{\alpha_2}z^{i_2}_{\alpha_2}
    S_{\alpha_3}z^{i_3}_{\alpha_3}
    S_{\alpha_4}z^{i_4}_{\alpha_4}.
    \label{eq:app:renormalization:renormalized-params-n-1-m-le-4}
\end{eqnarray}
\begin{eqnarray}
    \tilde{J}^{i_1,i_2}_{\nu_2;\alpha_1,\alpha_2}
    &=&
    C_2
    J^{i_1,i_2}_{\nu_2;\alpha_1,\alpha_2}
    +\nonumber\\&&+
    C_3
    \sum_{\substack{\nu_3,\\\alpha_3,\\i_3}}
    \Biggl(
        J^{i_1,i_2,i_3}_{\nu_2,\nu_3;\alpha_1,\alpha_2,\alpha_3}
        +\nonumber\\&&\phantom{+C_4\sum_{\substack{\nu_3,\\\alpha_3,\\i_3}}\Bigl(}+
        J^{i_1,i_3,i_2}_{\nu_3,\nu_2;\alpha_1,\alpha_3,\alpha_2}
        +\nonumber\\&&\phantom{+C_4\sum_{\substack{\nu_3,\\\alpha_3,\\i_3}}\Bigl(}+
        J^{i_3,i_1,i_2}_{-\nu_3,\nu_2-\nu_3;\alpha_3,\alpha_1,\alpha_2}
    \Biggr)
    S_{\alpha_3}z^{i_3}_{\alpha_3}
    +\nonumber\\&&+
    C_4
    \sum_{\substack{\nu_3,\nu_4\\\alpha_3,\alpha_4\\i_3,i_4}}
    \Biggl(
        J^{i_1,i_2,i_3,i_4}_{\nu_2,\nu_3,\nu_4;\alpha_1,\alpha_2,\alpha_3,\alpha_4}
        +\nonumber\\&&\phantom{+C_4\sum_{\substack{\nu_3,\nu_4\\\alpha_3,\alpha_4\\i_3,i_4}}\Bigl(}+
        J^{i_1,i_3,i_2,i_4}_{\nu_3,\nu_2,\nu_4;\alpha_1,\alpha_3,\alpha_2,\alpha_4}
        +\nonumber\\&&\phantom{+C_4\sum_{\substack{\nu_3,\nu_4\\\alpha_3,\alpha_4\\i_3,i_4}}\Bigl(}+
        J^{i_1,i_4,i_3,i_2}_{\nu_4,\nu_3,\nu_2;\alpha_1,\alpha_4,\alpha_3,\alpha_2}
        +\nonumber\\&&\phantom{+C_4\sum_{\substack{\nu_3,\nu_4\\\alpha_3,\alpha_4\\i_3,i_4}}\Bigl(}+
        J^{i_3,i_1,i_2,i_4}_{-\nu_3,\nu_2-\nu_3,\nu_4-\nu_3;\alpha_3,\alpha_1,\alpha_2,\alpha_4}
        +\nonumber\\&&\phantom{+C_4\sum_{\substack{\nu_3,\nu_4\\\alpha_3,\alpha_4\\i_3,i_4}}\Bigl(}+
        J^{i_4,i_1,i_3,i_2}_{-\nu_4,\nu_3-\nu_4,\nu_2-\nu_4;\alpha_4,\alpha_1,\alpha_3,\alpha_2}
        +\nonumber\\&&\phantom{+C_4\sum_{\substack{\nu_3,\nu_4\\\alpha_3,\alpha_4\\i_3,i_4}}\Bigl(}+
        J^{i_3,i_4,i_1,i_2}_{\nu_4-\nu_3,-\nu_3,\nu_2-\nu_3;\alpha_3,\alpha_4,\alpha_1,\alpha_2}
    \Biggr)
    S_{\alpha_3}z^{i_3}_{\alpha_3}
    S_{\alpha_4}z^{i_4}_{\alpha_4}.
    \label{eq:app:renormalization:renormalized-params-n-2-m-le-4}
\\
    \tilde{J}^{i_1,i_2,i_3}_{\nu_2,\nu_3;\alpha_1,\alpha_2,\alpha_3}
    &=&
    C_3
    J^{i_1,i_2,i_3}_{\nu_2,\nu_3;\alpha_1,\alpha_2,\alpha_3}
    +\nonumber\\&&+
    C_4
    \sum_{\nu_4,\alpha_4}
    \Biggl(
        J^{i_1,i_2,i_3,i_4}_{\nu_2,\nu_3,\nu_4;\alpha_1,\alpha_2,\alpha_3,\alpha_4}
        +\nonumber\\&&\phantom{+C_3\sum_{\nu_4,\alpha_4}\Biggl(}+
        J^{i_1,i_2,i_4,i_3}_{\nu_2,\nu_4,\nu_3;\alpha_1,\alpha_2,\alpha_4,\alpha_3}
        +\nonumber\\&&\phantom{+C_3\sum_{\nu_4,\alpha_4}\Biggl(}+
        J^{i_1,i_4,i_3,i_2}_{\nu_4,\nu_3,\nu_2;\alpha_1,\alpha_4,\alpha_3,\alpha_2}
        +\nonumber\\&&\phantom{+C_3\sum_{\nu_4,\alpha_4}\Biggl(}+
        J^{i_4,i_2,i_3,i_1}_{\nu_2-\nu_4,\nu_3-\nu_4,-\nu_4;\alpha_4,\alpha_2,\alpha_3,\alpha_1}
    \Biggr)
    S_{\alpha_4}
    z^{i_4}_{\alpha_4}.
    \label{eq:app:renormalization:renormalized-params-n-3-m-le-4}
\\
    \tilde{J}^{i_1,i_2,i_3,i_4}_{\nu_2,\nu_3,\nu_4;\alpha_1,\alpha_2,\alpha_3,\alpha_4}
    &=&
    C_4
    J^{i_1,i_2,i_3,i_4}_{\nu_2,\nu_3,\nu_4;\alpha_1,\alpha_2,\alpha_3,\alpha_4}.
    \label{eq:app:renormalization:renormalized-params-n-4-m-le-4}
\end{eqnarray}

Finally, we note that the unknown analytical form of the integer partition function is the main reason why the equation for the renormalized parameters can not be written in a simple way. The simplification (that lead to the Equation~\eqref{eq:app:renormalization:renormalized-parameters-all-diff} in a special case) relies on the hermiticity of the initial Hamiltonian and on the choice of symmetric gauge of the initial parameters. Both of those properties depend on the evaluation of the integer partition function. 

\subsection{Renormalized Hamiltonian}

The only part of the initial spin Hamiltonian that contributes to the bosonic LSWT Hamiltonian is
\begin{eqnarray}
    \mathcal{\hat{H}}
    \approx
    NE^{(0)}
    +
    \sum_{\mu, \alpha_1, i_1}
    \tilde{J}^{i_1}_{\alpha_1}
    \hat{\Delta}^{i_1}_{\mu, \alpha_1}
    +
    \sum_{\substack{\mu, \nu_2,\\ \alpha_1, \alpha_2,\\ i_1, i_2}}
    \tilde{J}^{i_1i_2}_{\nu_2; \alpha_1, \alpha_2}
    \hat{\Delta}^{i_1}_{\mu, \alpha_1}
    \hat{\Delta}^{i_2}_{\mu+\nu_2, \alpha_2}
    +
    \dots
    \label{eq:app:renormalization:renormalized_hamiltonian_first_two_terms}
\end{eqnarray}

The extension of all LSWT equations presented in this article for the $n>4$ terms of the initial Hamiltonian in Equation~\EqAllMightyHamiltonian of the main text can be done simply by including the new corrections into the renormalized parameters $\tilde{J}^{i_1}_{\alpha_1}$ and $\tilde{J}^{i_1,i_2}_{\nu_2;\alpha_1,\alpha_2}$. The rest of the LSWT equations would not change. This is the true advantage of the renormalization.

\newpage
\section{\label{app:one-operator-term}Extremum of classical energy}

The condition $O_{\alpha} = 0, \forall \alpha$ is equivalent to either $\tilde{J}_{\alpha}^i = 0, \forall \alpha, \forall i$ or $\bm{J}_{\alpha} || \bm{z}_{\alpha}, \forall \alpha$ (which follows directly from Equation~\EqOCoeff of the main text). Therefore, to prove that the extremum condition on $E^{(0)}$ is equivalent to $O_{\alpha} = 0, \forall \alpha$ we need to show that the extremum condition on $E^{(0)}$ is equivalent to either $J_{\alpha}^i = 0, \forall \alpha, \forall i$ or $\bm{J}_{\alpha} || \bm{z}_{\alpha}, \forall \alpha$.

First, we compute the first derivative of $E^{(0)}$, which is paramount to the following proof.

\begin{equation}
    \dfrac{\partial E^{(0)}}{\partial z_{\xi}^{j}}
    =
    S_{\xi}
    \sum_{n=1}^{\infty}
    C_n
    \sum_{\substack{\nu_{2},\dots,\nu_m,\\ \alpha_{2},\dots,\alpha_m,\\ i_{2},\dots,i_m}}
    \sum_{p_{1,n}}
    \Biggl(
        J^{p_{1,n}(j,i_2,\dots,i_m)}_{p_{1,n}(\nu_2,\dots,\nu_m;\xi,\alpha_2,\dots,\alpha_m)}
        \prod_{m=2}^n S_{\alpha_m}z^{i_m}_{\alpha_m}
    \Biggr)
    =
    S_{\xi}
    \tilde{J}_{\xi}^j.
     \label{eq:app:on-operator-term:E-first-derivative}
\end{equation}

Note that the first derivative of classical energy depends on the single-spin renormalized interaction parameter, that also appears in the equation for $O_{\alpha}$.

Next we consider two cases: when all $\bm{z}_{\alpha}$ are varied in 3D Euclidean space and when all $\bm{z}_{\alpha}$ are varied on a unit sphere.

\subsection{Case 1: variation in 3D space}

The extremum condition on $E^{(0)}$ reads as

\begin{equation}
    \dfrac{\partial E^{(0)}}{\partial z_{\xi}^{j}}
    =
    0,
    \qquad
    \forall \xi, \forall j.
    \label{eq:app:on-operator-term:extremum_condition_3D_cart}
\end{equation}
Which is equivalent to the condition $\tilde{J}_{\alpha}^i = 0, \forall \alpha, \forall i$ due to Equation~\eqref{eq:app:on-operator-term:E-first-derivative}. Therefore, extremum of $E^{(0)}$ is equivalent to $O_{\alpha} = 0, \forall \alpha$.

\subsection{Case 2: variation on a sphere}

In this case we consider a constrained optimization problem with the $M$ constrains 

\begin{equation}
    g_{\alpha}(\bm{z}_{\alpha})
    =
    |\bm{z}_{\alpha}|^2 - 1,
    \qquad \forall \alpha,
    \label{eq:app:one-operator-term:contrain_conditions}
\end{equation}
where $M$ is the amount of sites in the unit cell. Then, the Lagrangian is

\begin{equation}
    \mathcal{L}
    =
    E^{(0)}
    +
    \sum_{\alpha}
    \lambda_{\alpha}
    g_{\alpha}
\end{equation}
and the extremum condition reads as
\begin{equation}
    \dfrac{\partial \mathcal{L}}{\partial z_{\xi}^{j}}
    =
    \dfrac{\partial E^{(0)}}{\partial z_{\xi}^{j}}
    +
    \sum_{\alpha}
    \lambda_{\alpha}
    \dfrac{\partial g_{\alpha}}{\partial z_{\xi}^{j}}
    =
    S_{\xi}
    \tilde{J}_{\xi}^j
    +
    2
    \lambda_{\xi}
    z_{\xi}^j
    =
    0,
    \qquad \forall \xi, \forall j,
\end{equation}
which is equivalent to $\bm{J}_{\alpha} || \bm{z}_{\alpha}, \forall \alpha$. Therefore, extremum of $E^{(0)}$ is equivalent to $O_{\alpha} = 0, \forall \alpha$.

\newpage
\section{\label{app:quadratic-form}Grand-dynamical matrix}

Here we list the formulas for the elements of the quadratic form from Equation~\EqAllMightyHamiltonianUpToLswt that are implemented in Magnopy. The symmetric gauge is chosen for the initial and renormalized interaction parameters of the spin Hamiltonian as described in~\ref{app:gauge}.

The grand-dynamical $2M\times2M$ matrix is given by Equation~\EqGrandDynamicalMatrixForm of the main text, which we repeat below
\begin{equation}
    \bm{\mathcal{D}}(\bm{k})
    =
    \begin{pmatrix}
        \bm{D_1}(\bm{k}) & \bm{D_2}(\bm{k}) \\
        \overline{\bm{D_2}(-\bm{k})} & \overline{\bm{D_1}(-\bm{k})} \\
    \end{pmatrix},
    \label{eq:supp:grand-dynamical-matrix}
\end{equation}
where an overline indicates component-wise complex conjugation. Elements of the $M\times M$ matrices $\bm{D_1}(\bm{k})$ and $\bm{D_2}(\bm{k})$ are
\begin{eqnarray}
    (D_1)_{\alpha_1,\alpha_2}(\bm{k})
    &=&
    \dfrac{1}{2}
    \Biggl(
        \sqrt{S_{\alpha_1}S_{\alpha_2}}
        \sum_{i_1, i_2}
        \tilde{J}_{\alpha_1,\alpha_2}^{i_1,i_2}(\bm{k})
        p_{\alpha_1}^{i_1}
        \overline{p_{\alpha_2}^{i_2}}
        -
        [\alpha_1 = \alpha_2]
        \sum_{i_1}
        z_{\alpha_1}^{i_1}
        \tilde{J}_{\alpha_1}^{i_1}
    \Biggr),
    \nonumber\\
    (D_2)_{\alpha_1,\alpha_2}(\bm{k})
    &=&
    \dfrac{1}{2}
    \sqrt{S_{\alpha_1}S_{\alpha_2}}
    \sum_{i_1,i_2}
    \tilde{J}_{\alpha_1,\alpha_2}^{i_1,i_2}(\bm{k})
    p_{\alpha_1}^{i_1}
    p_{\alpha_2}^{i_2},
    \nonumber\\
    \tilde{J}_{\alpha_1,\alpha_2}^{i_1,i_2}(\bm{k})
    &=&
    \sum_{\nu_2}
    e^{i\bm{k}\cdot\bm{r}_{\nu_2}}
    \tilde{J}_{\nu_2;\alpha_1,\alpha_2}^{i_1,i_2},
    \label{eq:supp:grand-dynamical-matrix-parts}
\end{eqnarray}
where $[\alpha_1 = \alpha_2]$ is an Iverson bracket \cite{Iverson1962-us,Knuth1992}. If the initial Hamiltonian in Equation~\EqAllMightyHamiltonian of the main text is Hermitian then matrix $\bm{D_1}(\bm{k})$ is Hermitian and $\overline{\bm{D_2}(-\bm{k})} = \bm{D_2}^{\dagger}(\bm{k})$, which guarantees the hermiticity of the matrix $\bm{\mathcal{D}}(\bm{k})$. 

Alternatively, the matrix $\bm{\mathcal{D}}(\bm{k})$ can be expressed as
\begin{equation}
    \bm{\mathcal{D}}(\bm{k})
    =
    \bm{\mathcal{D}_0} + 
    \sum_{\nu_2}
    e^{i\bm{k}\cdot\bm{r}_{\nu_2}}
    \bm{\mathcal{D}_{\nu_2}},
\end{equation}
where 
\begin{equation}
    \bm{\mathcal{D}_{\nu_2}}
    =
    \begin{pmatrix}
        \bm{D_{1;\nu_2}} & \bm{D_{2;\nu_2}} \\
        \overline{\bm{D_{2;\nu_2}}} & \overline{\bm{D_{1;\nu_2}}} \\
    \end{pmatrix},
\end{equation}
\begin{equation}
    \bm{\mathcal{D}}
    =
    \begin{pmatrix}
        \bm{D_0} & 0 \\
        0 & \bm{D_0} \\
    \end{pmatrix}.
\end{equation}

Elements of the $M\times M$ matrices $\bm{D_0}$, $\bm{D_{1;\nu_2}}$, $\bm{D_{2;\nu_2}}$ are
\begin{eqnarray}
    \Bigl(D_{1;\nu_2}\Bigr)_{\alpha_1,\alpha_2}
    &=&
    \dfrac{1}{2}
    \sqrt{S_{\alpha_1}S_{\alpha_2}}
    \sum_{i_1, i_2}
    \tilde{J}_{\nu_2;\alpha_1,\alpha_2}^{i_1,i_2}
    p_{\alpha_1}^{i_1}
    \overline{p_{\alpha_2}^{i_2}},
    \nonumber\\
    \Bigl(D_{2;\nu_2}\Bigr)_{\alpha_1,\alpha_2}
    &=&
    \dfrac{1}{2}
    \sqrt{S_{\alpha_1}S_{\alpha_2}}
    \sum_{i_1,i_2}
    \tilde{J}_{\nu_2;\alpha_1,\alpha_2}^{i_1,i_2}
    p_{\alpha_1}^{i_1}
    p_{\alpha_2}^{i_2},
    \nonumber\\
    (D_0)_{\alpha_1,\alpha_2}
    &=&
    -\dfrac{1}{2}
    [\alpha_1 =\alpha_2]
    \sum_{i_1}
    z_{\alpha_1}^{i_1}
    \tilde{J}_{\alpha_1}^{i_1}.
\end{eqnarray}
If the initial Hamiltonian in Equation~\EqAllMightyHamiltonian of the main text is Hermitian then $\bm{D_0}^{\dagger} = \bm{D_0}$, $\bm{D_{1;\nu_2}} = \bm{D^{\dagger}_{1;-\nu_2}}$ and $\bm{D_{2;\nu_2}} = \bm{D^T_{2;-\nu_2}}$, which guarantees the hermiticity of $\bm{\mathcal{D}}(\bm{k})$.

\subsection{Inversion symmetry}

The LSWT Hamiltonian satisfies inversion symmetry as defined in \cite{Ohashi2020} when $\tilde{J}_{\alpha_1,\alpha_2}^{i_1,i_2}(\bm{k}) = \tilde{J}_{\alpha_1,\alpha_2}^{i_1,i_2}(-\bm{k})$. In particular, the inversion symmetry is satisfied when 

\begin{equation}
    \tilde{J}_{\nu_2;\alpha_1,\alpha_2}^{i_1,i_2}
    =
    \tilde{J}_{\nu_2;\alpha_2,\alpha_1}^{i_2,i_1},
    \qquad
    \forall \nu_2,\forall \alpha_1, \forall \alpha_2.
    \label{eq:app:quadratic-form:condition_of_inversion_symmetry}
\end{equation}

Note, that Equation~\eqref{eq:app:quadratic-form:condition_of_inversion_symmetry} is different from the implication of hermiticity (\ref{app:hermiticity}) combined with the choice of symmetric gauge (\ref{app:gauge}) for $n = 2$, that reads as $\tilde{J}_{\nu_2;\alpha_1,\alpha_2}^{i_1,i_2}=\tilde{J}_{-\nu_2;\alpha_2,\alpha_1}^{i_2,i_1}$

\newpage
\section{\label{app:lswt-diagonalisation}Diagonalization of quadratic Bogoliubov Hamiltonian}

Here we describe the details of the Colpa's method \cite{Colpa1978} to diagonalize the LSWT bosonic Hamiltonian and give a proof of Equation~\EqLinkEnergiesPlusMinusK~and~\EqLinkGPlusMinusK of the main text. The latter is important for reducing computational effort when performing a calculation with a large number of k-points evenly distributed across the entire Brillouin zone.

The starting point is the Bogoliubov bosonic Hamiltonian (left-hand and central parts of Equation~\EqBilinearPartBogoliubovHamiltonian of the main text)
\begin{equation}
    \mathcal{\hat{H}}^{B}
    =
    \sum_{\bm{k}}
    \bm{\mathcal{\hat{A}}}^{\dagger}(\bm{k})
    \bm{\mathcal{D}}(\bm{k})
    \bm{\mathcal{\hat{A}}}(\bm{k})
\end{equation}
where $\bm{\mathcal{D}}(\bm{k})$ is defined in Equation~\EqGrandDynamicalMatrixForm and $\bm{\mathcal{\hat{A}}}(\bm{k})$ is defined in Equation~\EqVectorBosonsBeforeDiagonalization of the main text. We define the Bogoliubov transformation, that diagonalizes the Hamiltonian, as
\begin{equation}
    \bm{\hat{B}}(\bm{k})
    =
    \bm{\mathcal{G}}(\bm{k})\bm{\mathcal{\hat{A}}}(\bm{k})
    \label{eq:app:lswt-diagonalisation:Bogoliubol-transformation}
\end{equation}
where 
$\bm{\mathcal{G}}(\bm{k})$ is the $2M\times2M$ para-unitary transformation matrix, and the new boson operators are arranged into the column vector as
\begin{eqnarray}
    \bm{\mathcal{\hat{B}}}(\bm{k}) &=& \Bigl(\hat{b}_{1}(\bm{k}), \dots, \hat{b}_{M}(\bm{k}), \hat{b}_{1}^{\dagger}(-\bm{k}), \dots, \hat{b}_{M}^{\dagger}(-\bm{k})\Bigr)^T
\end{eqnarray}
Altogether, the diagonalized Hamiltonian is written as a quadratic form
\begin{eqnarray}
    \mathcal{\hat{H}}^{B}
    = 
    \sum_{\bm{k}}
    \bm{\mathcal{\hat{B}}}^{\dagger}(\bm{k})
    \bm{\mathcal{E}}(\bm{k})
    \bm{\mathcal{\hat{B}}}(\bm{k})
\end{eqnarray}
where all elements of the $2 M\times 2 M$ diagonal matrix 
\begin{equation}
    \bm{\mathcal{E}}(\bm{k})
    =
    \Bigl(\bm{\mathcal{G}}^\dagger(\bm{k})\Bigr)^{-1}
    \bm{\mathcal{D}}(\bm{k})
    \Bigl(\bm{\mathcal{G}}(\bm{k})\Bigr)^{-1}
    \label{eq:app:lswt-diagonalisation:eigenenergies-matrix-transformation}
\end{equation}
are real and positive. We note that the Hamiltonian can be diagonalized and has positive real eigenvalues only if the matrix $\bm{\mathcal{D}}(\bm{k})$ is positive-definite \cite{Colpa1978}. Non-positive definiteness, which can arise in numerical calculations, indicates that the assumed vacuum state $\ket{0}$ is unstable. 

In the next two sections we demonstrate the connection of the eigenenergies ($\bm{\mathcal{E}}(\bm{k})$ and $\bm{\mathcal{E}}(-\bm{k})$) and eigenstates ($\bm{\mathcal{G}}(\bm{k})$ and $\bm{\mathcal{G}}(-\bm{k})$) for positive and negative crystal quasi-momentum vectors $\bm{k}$ and $-\bm{k}$ based on the results presented by Colpa in Ref. \cite{Colpa1978} and on Equation~\EqGDMPlusMinusK of the main text.

\subsection{Eigenenergies}

In this subsection we prove Equation~\EqLinkEnergiesPlusMinusK of the main text. We start from Equation~2.14 of \cite{Colpa1978} written for positive and negative crystal quasi-momenta
\begin{subequations}
\begin{equation}
    \det\Bigl(\bm{\mathcal{D}}(\bm{k}) - \lambda_{\rho}(\bm{k})\bm{g}\Bigr) = 0
    \label{eq:app:lswt-diagonalisation:secular-plus-k}
\end{equation}
\begin{equation}
    \det\Bigl(\bm{\mathcal{D}}(-\bm{k}) - \lambda_{\rho}(-\bm{k})\bm{g}\Bigr) = 0
    \label{eq:app:lswt-diagonalisation:secular-minus-k}
\end{equation}
\label{eq:app:lswt-diagonalisation:secular}
\end{subequations}
where $\lambda_{\rho}$ are the diagonal elements of the diagonal matrix $\bm{\mathcal{L}}(\bm{k}) = \bm{g}\,\bm{\mathcal{E}}(\bm{k})$. Index $\rho$ runs from $1$ to $2M$ (on contrary to the indices $\alpha$ or $\beta$ that run from $1$ to $M$) and $\bm{g}$ is a $2M \times 2M$ para-unit matrix defined as
\begin{equation}
    \bm{g}
    =
    \sigma_3 \otimes \bm{I}
    =
    \begin{pmatrix}
        \bm{I} &  \bm{0} \\
        \bm{0} & -\bm{I} \\
    \end{pmatrix},
\end{equation}
where $\sigma_3$ is the third $2\times2$ Pauli matrix and $\bm{I}$ is an $M\times M$ identity matrix. Equation~\eqref{eq:app:lswt-diagonalisation:secular} are the secular equations for positive and negative crystal quasi-momenta, whose solutions define the eigenenergies of the normal magnon modes. By multiplying Equation~\eqref{eq:app:lswt-diagonalisation:secular-minus-k} by $\bm{\mathcal{I}}$ (defined in Equation~\EqGDMPlusMinusK of the main text) on the left and on the right, taking a complex conjugate and using Equation~\EqGDMPlusMinusK of the main text (note that $\bm{\mathcal{I}}\bm{g}\bm{\mathcal{I}} = -\bm{g}$) we obtain
\begin{equation}
    \det\Bigl(\bm{\mathcal{D}}(\bm{k}) + \overline{\lambda_{\rho}(-\bm{k})}\bm{g}\Bigr)
    =
    0
    \label{eq:app:lswt-diagonalisation:secular-minus-k-tranformed}
\end{equation}

Comparing Equation~\eqref{eq:app:lswt-diagonalisation:secular-minus-k-tranformed} with Equation~\eqref{eq:app:lswt-diagonalisation:secular-plus-k} we conclude that there exist $2M$ distinct pairs of $\rho_1, \rho_2$ such that $\lambda_{\rho_1}(\bm{k}) = -\overline{\lambda_{\rho_2}(-\bm{k})}$. Diagonal elements of $\bm{\mathcal{L}}(\bm{k})$ can always be arranged in a the way that the first $M$ elements are positive and the last $M$ elements are negative \cite{Colpa1978}, and order of elements within each group is arbitrary. Therefore, we can choose that order such as
\begin{eqnarray}
    \lambda_{\beta}(\bm{k}) &=& -\overline{\lambda_{M+\beta}(-\bm{k})} \nonumber\\
    \lambda_{M+\beta}(\bm{k}) &=& -\overline{\lambda_{\beta}(-\bm{k})}
    \label{eq:app:lswt-diagonalisation:link-L-plus-minus-k}
\end{eqnarray}
where $\beta$ is the index that enumerates normal magnon modes and runs from $1$ to $M$. The complex conjugation is redundant as all elements of $\bm{\mathcal{L}}(\bm{k})$ are real if the grand-dynamical matrix is Hermitian and positive-definite \cite{Colpa1978}. Equation~\EqLinkEnergiesPlusMinusK of the main text directly follows from Equation~\eqref{eq:app:lswt-diagonalisation:link-L-plus-minus-k}.

\subsection{Transformation matrix}

In this subsection we prove Equation~\EqLinkGPlusMinusK of the main text. We start from Equation~2.13 of \cite{Colpa1978} written for positive and negative crystal quasi-momenta 
\begin{subequations}
\begin{equation}
    \Bigl(
        \bm{\mathcal{D}}(\bm{k})
        -
        \lambda_{\rho}(\bm{k})
        \bm{g}
    \Bigr)
    \bm{w_{\rho}(\bm{k})}
    =
    0
    \label{eq:app:lswt-diagonalisation:eigenvector-eq-plus-k}
\end{equation}
\begin{equation}
    \Bigl(
        \bm{\mathcal{D}}(-\bm{k})
        -
        \lambda_{\rho}(-\bm{k})
        \bm{g}
    \Bigr)
    \bm{w_{\rho}(-\bm{k})}
    =
    0
    \label{eq:app:lswt-diagonalisation:eigenvector-eq-minus-k}
\end{equation}
    \label{eq:app:lswt-diagonalisation:eigenvector-eq}
\end{subequations}
where
\begin{equation}
    \bm{w_{\rho}(\bm{k})}
    =
    \begin{pmatrix}
        \bm{v}_{\rho}(\bm{k}) \\
        \bm{u}_{\rho}(\bm{k}) 
    \end{pmatrix}
\end{equation}
is a $\rho^{\text{th}}$ column of the inverse of the transformation matrix $\Bigl(\bm{\mathcal{G}}(\bm{k})\Bigr)^{-1}$. By multiplying Equation~\eqref{eq:app:lswt-diagonalisation:eigenvector-eq-minus-k} by $\bm{\mathcal{I}}$ on the left, taking a complex conjugate, using Equation~\EqGDMPlusMinusK of the main text, and Equation~\eqref{eq:app:lswt-diagonalisation:link-L-plus-minus-k}, we get
\begin{equation}
    \Bigl(
        \bm{\mathcal{D}}(\bm{k})
        -
        \lambda_{\rho}(\bm{k})
        \bm{g}
    \Bigr)
    \bm{\mathcal{I}}
    \overline{\bm{w_{\rho}(-\bm{k})}}
    =
    0
    \label{eq:app:lswt-diagonalisation:eigenvector-eq-minus-k-transform}
\end{equation}

By comparing Equation~\eqref{eq:app:lswt-diagonalisation:eigenvector-eq-minus-k-transform} with Equation~\eqref{eq:app:lswt-diagonalisation:eigenvector-eq-plus-k} we conclude that there are $2M$ pairs of $\rho_1, \rho_2$ such that
\begin{eqnarray}
    \bm{u}_{\rho_1}(\bm{k}) &=& \overline{\bm{v}_{\rho_2}(-\bm{k})} \nonumber\\
    \bm{v}_{\rho_1}(\bm{k}) &=& \overline{\bm{u}_{\rho_2}(-\bm{k})} 
    \label{eq:app:lswt-diagonalisation:link-u-v-plus-minus-k}
\end{eqnarray}

Equation~\eqref{eq:app:lswt-diagonalisation:link-u-v-plus-minus-k},~\eqref{eq:app:lswt-diagonalisation:link-L-plus-minus-k} together with Equation~3.7 of \cite{Colpa1978} directly lead to the Equation~\EqLinkGPlusMinusK of the main text.

\subsection{Serialization of a matrix form}

Finally, we provide some details about the transition from Equation~\EqBilinearPartBogoliubovHamiltonian to Equation~\EqMatrixToSerialDiagonalized of the main text. First, we expand the matrix form and rewrite the sum over $\bm{k}$ as
\begin{eqnarray}
    \mathcal{\hat{H}}^{B}
    &=&
    \sum_{\bm{k}, \beta}
    \Biggl(
        \tilde{\omega}_{\beta}(\bm{k})
        \hat{b}^{\dagger}_{\beta}(\bm{k})\hat{b}_{\beta}(\bm{k})
        +
        \tilde{\omega}_{M+\beta}(\bm{k})
        \hat{b}_{\beta}(-\bm{k})\hat{b}^{\dagger}_{\beta}(-\bm{k})
    \Biggr)
    \nonumber=\\&=&
    \sum_{\bm{k} > 0, \beta}
    \Biggl(
        \tilde{\omega}_{\beta}(\bm{k})
        \hat{b}^{\dagger}_{\beta}(\bm{k})\hat{b}_{\beta}(\bm{k})
        +
        \tilde{\omega}_{M+\beta}(\bm{k})
        \hat{b}_{\beta}(-\bm{k})\hat{b}^{\dagger}_{\beta}(-\bm{k})
        +\nonumber\\&&\phantom{\sum_{\bm{k} > 0, \beta}\Biggl(}+
        \tilde{\omega}_{\beta}(-\bm{k})
        \hat{b}^{\dagger}_{\beta}(-\bm{k})\hat{b}_{\beta}(-\bm{k})
        +
        \tilde{\omega}_{M+\beta}(-\bm{k})
        \hat{b}_{\beta}(\bm{k})\hat{b}^{\dagger}_{\beta}(\bm{k})
    \Biggr),
\end{eqnarray}
then we apply Equation~\EqLinkEnergiesPlusMinusK of the main text and use the bosonic commutation relation
\begin{eqnarray}
    \mathcal{\hat{H}}^{B}
    &=&
    \sum_{\bm{k} > 0, \beta}
    \Biggl(
        \tilde{\omega}_{\beta}(\bm{k})
        \hat{b}^{\dagger}_{\beta}(\bm{k})\hat{b}_{\beta}(\bm{k})
        +
        \tilde{\omega}_{\beta}(-\bm{k})
        \Bigl(\hat{b}^{\dagger}_{\beta}(-\bm{k})\hat{b}_{\beta}(-\bm{k}) + 1\Bigr)
        +\nonumber\\&&\phantom{\sum_{\bm{k} > 0, \beta}\Biggl(}+
        \tilde{\omega}_{\beta}(-\bm{k})
        \hat{b}^{\dagger}_{\beta}(-\bm{k})\hat{b}_{\beta}(-\bm{k})
        +
        \tilde{\omega}_{\beta}(\bm{k})
        \Bigl(\hat{b}^{\dagger}_{\beta}(\bm{k})\hat{b}_{\beta}(\bm{k}) + 1\Bigr)
    \Biggr)
    \nonumber=\\&=&
    \sum_{\bm{k} > 0, \beta}
    \Biggl(
        2\tilde{\omega}_{\beta}(\bm{k})
        \Bigl(\hat{b}^{\dagger}_{\beta}(\bm{k})\hat{b}_{\beta}(\bm{k}) + \dfrac{1}{2}\Bigr)
        +
        2\tilde{\omega}_{\beta}(-\bm{k})
        \Bigl(\hat{b}^{\dagger}_{\beta}(-\bm{k})\hat{b}_{\beta}(-\bm{k}) + \dfrac{1}{2}\Bigr)
    \Biggr)
    \nonumber=\\&=&
    \sum_{\bm{k}, \beta}
    2\tilde{\omega}_{\beta}(\bm{k})
    \Biggl(
        \hat{b}^{\dagger}_{\beta}(\bm{k})\hat{b}_{\beta}(\bm{k}) + \dfrac{1}{2}
    \Biggr),
\end{eqnarray}
which leads directly to the the Equation~\EqMatrixToSerialDiagonalized of the main text.

\newpage
\section{\label{app:analytical-models}Analytical models}
Here we provide a formal definition of the two model Hamiltonians mentioned in the main text. These two models are proposed as a bare minimum set of standard analytical magnon problems for testing any numerical implementation of LSWT. Both models are defined in the same global reference frame $(\bm{x}, \bm{y}, \bm{z})$ as depicted in Figure~\FigModelSAB of the main text.

\subsection{Model 1: ferromagnet}

A ferromagnet with the four nearest neighbors on a square lattice. The solution of this model is given by Equation~\EqModelOneOmegas of the main text.

\begin{eqnarray}
    \mathcal{\hat{H}}
    &=&
    g\mu_B
    \bm{B}
    \cdot
    \sum_{i}
    \bm{\hat{S}}_i
    -
    \sum_i
    \bm{\hat{S}}_i
    \cdot
    \bm{A}_i
    \cdot
    \bm{\hat{S}}_i
    +
    \sum_{i,j}
    J_{i,j}\,
    \bm{\hat{S}}_i
    \cdot
    \bm{\hat{S}}_j
    +
    \sum_{i,j}
    \bm{D}_{ij}
    \cdot
    \Bigl(
    \bm{\hat{S}}_i
    \times
    \bm{\hat{S}}_j
    \Bigr),
\end{eqnarray}
where $J < 0$, $A>0$, $\bra{0}\hat{S}_i^{z_i}\ket{0} = S$, $g=2$ is a g-factor, $\mu_B$ is the Bohr magneton, $|\bm{D}_{ij}|$ is small enough for the ground state to be ferromagnetic, $\bm{z}_i = (\bm{x}-\bm{y})/\sqrt{2}$, and
\begin{equation}
    \bm{A}_i
    =
    \begin{pmatrix}
        A/2  & -A/2 & 0 \\
        -A/2 &  A/2 & 0 \\
           0 &    0 & 0 \\
    \end{pmatrix}.
    \label{eq:app:analytical-models:on_site_anisotropy}
\end{equation}

Double counting is implied in the sum (i.~e. both $i\rightarrow j$ and $j \rightarrow i$ are included). $J_{ij}$ and $\bm{D}_{ij}$ are defined as in the Figure~\FigModelSA. If $B = 0$, $A = 0$ and $D = 0$, then this is a textbook problem \cite{blundell2001magnetism,coey2010magnetism,gurevich2020magnetization,jensen1991rare,rezende2020fundamentals,simon2013oxford,white2007quantum}.

\subsection{Model 2: antiferromagnet}

An antiferromagnet with the four nearest neighbors on a cubic lattice. The solution of this model is given by Equation~\EqModelTwoOmegas of the main text.

\begin{equation}
    \mathcal{\hat{H}}
    =
    g\mu_B
    \bm{B}
    \cdot
    \sum_{i}
    \bm{\hat{S}}_i
    -
    \sum_i
    \bm{\hat{S}}_i
    \cdot
    \bm{A}_i
    \cdot
    \bm{\hat{S}}_i
    +
    \sum_{i,j}
    J_{i,j}\,
    \bm{\hat{S}}_i
    \cdot
    \bm{\hat{S}}_j,
\end{equation}
where $J > 0$, $A>0$, $\bra{0}\hat{S}_i^{z_i}\ket{0} = S$, $g=2$ is a g-factor, $\mu_B$ is the Bohr magneton, $|\bm{B}|$ is small enough for the ground state to be antiferromagnetic, $\bm{z}_i = (\bm{x}-\bm{y})/\sqrt{2}$ for red (i.~e. pointing to the bottom right corner) sites in Figure~\FigModelSB, $\bm{z}_i = (-\bm{x}+\bm{y})/\sqrt{2}$ for blue (i.~e. pointing to the upper left corner) sites in Figure~\FigModelSB, and matrix $\bm{A}_i$ is the same as in Equation~\eqref{eq:app:analytical-models:on_site_anisotropy}. Double counting is implied in the sum (i.~e. both $i\rightarrow j$ and $j \rightarrow i$ are included). $J_{ij}$ are defined as in the Figure~\FigModelSB.

\end{document}